\documentclass[a4,12pt]{article}
\usepackage{amssymb}
 \usepackage{latexsym}
\usepackage{epsfig}
 \usepackage{graphicx}
\usepackage{psfrag}
     
\newcommand{\R}{{\mathbb R}}
 \newcommand{\D}{{\cal D}}
\newcommand{\Sbb}{\mathbb S}

 \newcommand{\C}{{\mathbb C}}
\newcommand{\N}{{\mathbb N}}

 \def\hat{\widehat}
\def\tilde{\widetilde}
 \def\bfo{\begin{eqnarray*} }
\def\efo{\end{eqnarray*} }
 \def\ba{\begin{eqnarray*} }
\def\ea{\end{eqnarray*} }
 \def\beq{\begin{eqnarray}}
\def\eeq{\end{eqnarray}}
 \def\supp{\hbox{supp}\,}

 \def\dist{\hbox{dist}}

 \def\det{\hbox{det}\,}
\def\bra{\langle}
 
\def\cet{\rangle}

 \def\e{\varepsilon}
\def\p{\partial}
 \def\a{\alpha}

\def\tE{\tilde E}
 
\def\M{{\mathcal M}}
 \def\F{{\mathcal F}}

\def\stwo{\Sbb^2}

\def\ie{\emph{i.e.}}
 \def\eg{\emph{e.g.}}
\def\R{\mathbb R}
 \def\bq{\begin{equation}}
\def\eq{\end{equation}}
 \def\D{\Delta}
\def\bE{{\bf{E}}}
 \def\bH{{\bf{H}}}
\def\te{\tilde{\e}}
 \def\tm{\tilde{\mu}}
\def\lra{\longrightarrow}
 \def\tH{\tilde{H}}

\def\ni{\noindent}

\def\num#1{\medskip\ni{\bf (#1)\quad}}

 \newtheorem{definition}{Definition}[section]
\newtheorem{theorem}[definition]{Theorem}
 \newtheorem{lemma}[definition]{Lemma}

 \newtheorem{remark}[definition]{Remark}

 \def\tg{\tilde{g}}
\def\tr{\tilde\rho}
 \def\tu{\tilde{u}}
\def\N{{\mathcal N}}
 \def\g{\gamma}

\begin{document}

 \title{Invisibility and Inverse Problems}

\author{Allan Greenleaf\footnote{Department of Mathematics,
 University of
Rochester, Rochester, NY 14627, USA. Partially
 supported by NSF grant
DMS-0551894. }\\
 Yaroslav Kurylev\footnote{Department of Mathematics,
University College London,
 Gower Street, London, WC1E 5BT, UK}
\\
 Matti Lassas,\footnote{
Helsinki University of Technology, Institute of Mathematics,
 P.O.Box 1100, FIN-02015, Finland. Partially supported by Academy of
Finland CoE Project 213476.}
 \\
Gunther Uhlmann\footnote{Department of Mathematics,
 University of Washington, Seattle, WA 98195, USA. Partially supported
by the NSF and a Walker
 Family Endowed Professorship.}
}

\date{}
 \maketitle

 \begin{abstract}

We  describe recent theoretical and experimental progress on
 making objects invisible. 
Ideas for devices that would have once seemed fanciful  may now be
 at least approximately realized physically,   using a new class of
artificially structured materials, \emph{metamaterials}.
 The equations that govern a variety of wave phenomena, including 
 electrostatics, electromagnetism, acoustics and quantum mechanics,
 have transformation laws under changes of variables which allow one to design 
material parameters that
 steer waves around a hidden region, returning them to their
original path on the far side. Not only are observers 
 unaware of the contents of the hidden region, they are not even 
aware that something is being hidden; the object, which 
 casts no shadow, is said to be \emph{cloaked}.  Proposals
for,  and even experimental  implementations of, such cloaking devices  have
 received the most attention, but other devices
having striking effects on wave propagation,
 unseen in nature,
 are also possible. These
 designs are initially based on the
transformation laws of the relevant  PDEs, but
due to
the singular transformations needed for the desired effects, care needs to
 be
taken in formulating and analyzing physically meaningful solutions. We
 recount
the recent history of the subject and discuss some of the mathematical and
 physical issues involved.

 \end{abstract}

 \section{Introduction}\label{sec-intro}

 Invisibility has been a subject of human fascination for millennia,
from the Greek legend of  Perseus versus Medusa to the more recent \emph{The
 Invisible Man} and  \emph{Harry Potter}. Over the years, there
have been occasional scientific prescriptions for invisibility in various
 settings, \eg, \cite{Ke,Ben}.  
 However, since 2005 there has been a wave of
serious theoretical proposals
 \cite{AE,MN,Mil,Le,PSS1} in the physics literature,
{and a widely reported experiment by Schurig et al.
 \cite{Sc},} for
cloaking devices -- structures that would not only make an object
 invisible
but also undetectable to electromagnetic waves, thus making it \emph{cloaked}.
The particular  route to
 cloaking that has received the most attention is that of {\sl
 transformation
optics} \cite{ward}, the design of optical devices with customized
 effects  on wave
propagation, made possible by taking advantage of the transformation rules
 for
the material properties of optics: the index of refraction $n(x)$ for scalar
 optics, governed by the Helmholtz equation, and the electrical permittivity
$\e(x)$ and magnetic permeability
 $\mu(x)$ for vector optics, as described by Maxwell's equations.
It is this approach to cloaking, and other novel effects on wave
propagation, that we will examine here.

 As it happens, two papers
appeared in the same 2006 issue of \emph{Science} with transformation
 optics-based proposals for cloaking. Leonhardt \cite{Le} gave a
 description,
based on conformal mapping,  of inhomogeneous indices of refraction $n$  in
 two dimensions that would cause light rays to go around a  region and
emerge on the other side as if they had passed through empty space (for
 which $n(x)\equiv 1$).  On the other hand, Pendry, Schurig and
 Smith \cite{PSS1} gave a prescription for values of $\e$ and $\mu$ yielding a
cloaking device for electromagnetic waves, based on the fact that $\e$ and
 $\mu$ transform nicely under changes of variables, cf. (\ref{eqn1.6}).
In fact, this construction used  the same singular
transformation (\ref{eqn-sing transf0}) 
as had  been used
three years earlier \cite{GLU2,GLU3} to describe examples of
nondetectability in the context of   Calder\'on's  Problem for conductivity,
which transforms in the same way as $\e$ and $\mu$.

We  briefly outline here the basic ideas of transformation optics, in the context of electrostatics,  leading  to a theoretical blueprint of a conductivity that cloaks an object from  observation using electrostatic measurements  \cite{GLU2,GLU3}. Given that the invariance of the underlying equation is a crucial ingredient of transformation optics it is natural
to set Calder\'on's problem on a compact Riemannian manifold with boundary,
$(M,g)$ with $g$ the Riemannian metric and boundary $\p M$ where the observations are made.
The
Laplace-Beltrami operator associated to  $g$ is given in
local coordinates by
\begin{equation}\label{eqn1.8}
\Delta_ g u=\frac{1}{\sqrt{|g|}}\sum^n_{i,j=1}\frac{\p}{\p x_i}
\left(\sqrt{|g|} g^{ij} \frac{\p u}{\p x_j}\right)
\end{equation}
where $(g^{ij})$ is the matrix inverse of the metric tensor $(g_{ij})$
and $|g|=\det g$.
 Let
us consider the Dirichlet problem associated to (\ref{eqn1.8}),
\begin{equation}\label{eqn1.9}
\Delta_ g u  = 0\hbox{ on } M,\quad u|_{\partial M}=f.
\end{equation}
We define the Dirichlet-to-Neumann (DN) map in this case by
\beq\label{eqn1.10}
\Lambda_g (f)=\sum^n_{i,j=1}\left.\left(\nu_i g^{ij}\sqrt{|g|} \frac {\partial u} {\partial x_j}\right)
\right|_{\partial M}\quad 
\eeq
where $\nu$ denotes the unit outer normal.
Calder\'on's (inverse) problem, the question of whether one can recover $g$ from $\Lambda_g$, has been the subject of a tremendous amount of work over the last quarter century. In Sec.\ 2, we briefly summarize the history and current status of this problem.

Given the invariant formulation of the DN map, it is straightforward to see that

\begin{equation}\label{eqn1.11}
\Lambda_{\psi^\ast g} = \Lambda_g
\end{equation}
for any $C^\infty$ diffeomorphism $\psi$  of $\overline M$ which
is the identity on the boundary. As usual, $\psi^\ast g$ denotes the
pull back of the metric $g$ by the diffeomorphism $\psi$.
For domains in Euclidean space of dimension $n\ge 3$, 
the metric $g$ corresponds to an anisotropic
conductivity $\sigma$, represented by the symmetric matrix-valued function 

\begin{equation}\label{conductivity}
\sigma^{ij}=|g|^{1/2} g^{ij}.
\end{equation}
The DN map sends the voltage potential at the boundary to the induced current flux.

The invariance (\ref{eqn1.11}) can be considered as a weak form of invisibility.
However, although the (generally distinct) media represented by $g$ and $\psi^* g$
are indistinguishable by boundary observations,
 nothing has yet been hidden.
In cloaking,  we are looking for a 
way to hide from  boundary measurements both  \emph{an object} 
 enclosed in some domain $D$ and \emph{the fact} that it is being hidden.
Suppose now that an object we want to cloak is enclosed in the ball of radius one, $B(0,1)$, and that we measure the DN map on the boundary of the the ball of radius two, $B(0,2)$.
Motivated by degenerations of singular Riemannian manifolds (see Sec.\ 3)
consider the
following \emph{singular} transformation stretching (or ``blowing up'') the origin to the ball $\overline{B}(0,1)$: 

\beq
\label{eqn-sing transf0}
& &F_1:B(0,2)\setminus\{0\}\to
B(0,2)\setminus\overline B(0,1),\\ \nonumber
 & & F_1(x)=(\frac {|x|}2+1)\frac x{|x|},\quad
0<|x|<2.
\eeq

\begin{figure}[htbp]\label{figure for map F.}
\psfrag{1}{$F_1$}
\begin{center}
\includegraphics[width=.7\linewidth]{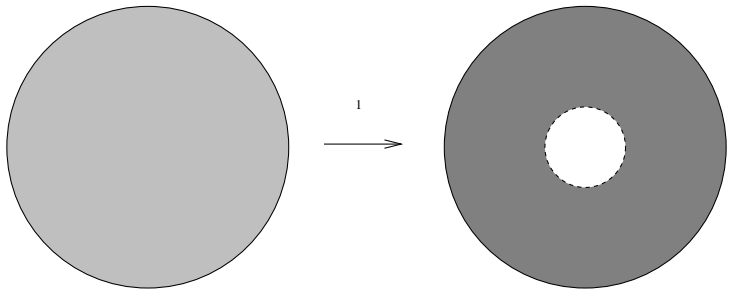}
\caption{Map $F_1:B(0,2)\setminus\{0\}\to B(0,2)\setminus\overline B(0,1)$}
\end{center}
\end{figure}

Also note that the metric $\tilde{g}=(F_1)_{*} g_0$, where $(F_1)_*= (F_1^{-1})^*$ and $g_0$ is the Euclidean metric, is
singular on the unit sphere $\mathbb S^{n-1}$, the interface between the cloaked and uncloaked regions, which we call the \emph{cloaking surface}. 
In fact, the conductivity $\tilde\sigma$ associated to this 
metric by (\ref{conductivity}) has zero  and/or infinite eigenvalues (depending on the dimension) as $r\searrow 1$. In $\R^3$, $\tilde\sigma$ is given in
spherical coordinates 
$(r,\phi,\theta)\mapsto
      (r\sin\theta \cos \phi,r\sin\theta \sin \phi,r\cos\theta)$ 
by

\begin{equation}\label{eqn-sing tensor 1} 
\tilde \sigma=
\left(\begin{array}{ccc}
2(r-1)^2\sin \theta & 0 & 0\\
0 & 2 \sin \theta & 0 \\
0 & 0 &  2 (\sin \theta)^{-1}\\
\end{array}
\right), \quad 1<|x|\leq 2.
\end{equation}
Note that  $\tilde \sigma$ is singular
(degenerate)
on
the sphere of radius $1$ in the sense that it is not bounded from below by any positive
multiple of the identity matrix $I$. (See \cite{KSVW} for a similar calculation.)

\begin{figure}[htbp]\label{Analytic sol.}
\begin{center}
\includegraphics[width=.7\linewidth]{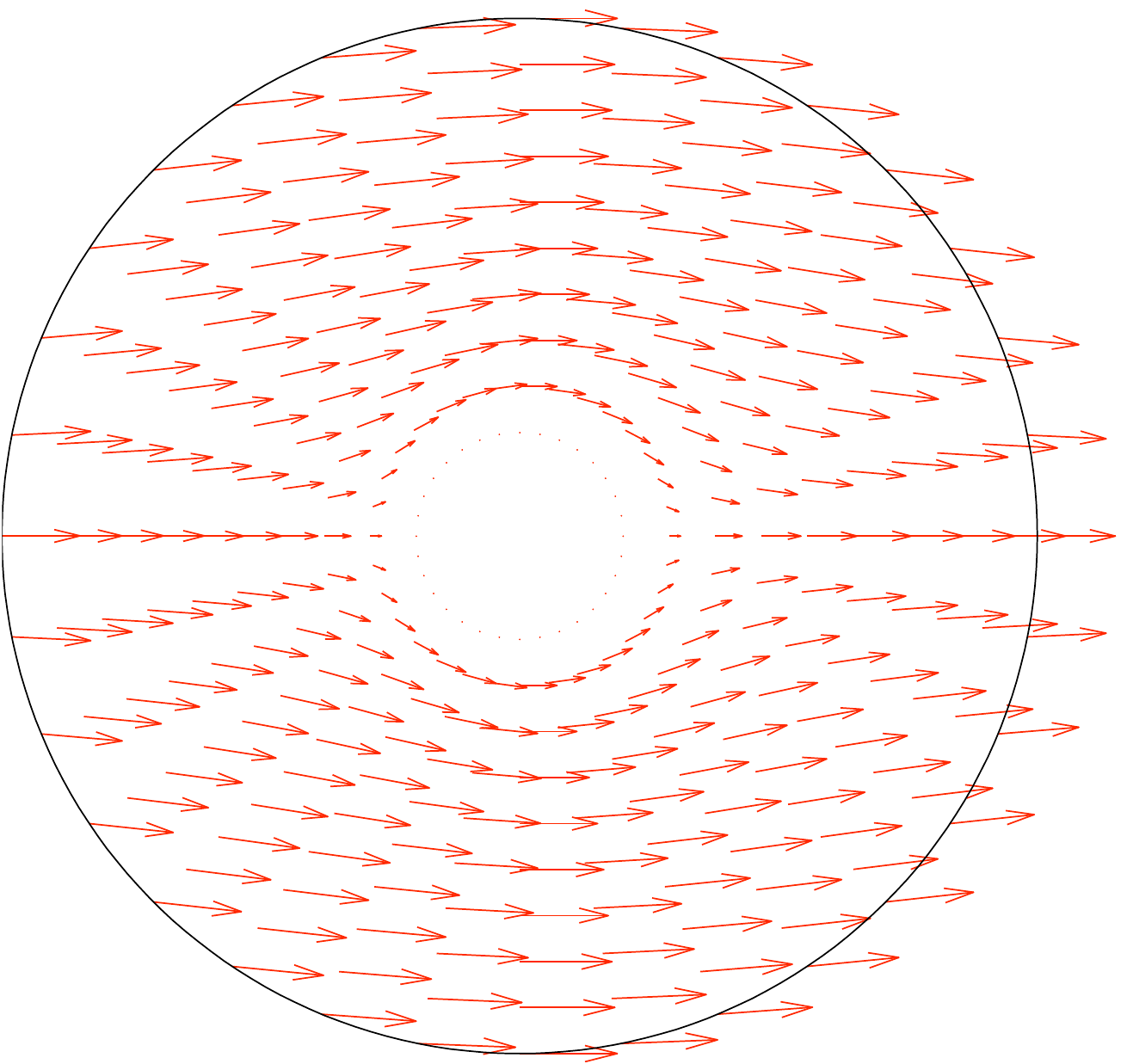}
\caption{Analytic solutions for the currents}
\end{center}
\end{figure}

The currents associated to this singular conductivity on $B(0,2)\setminus \overline{B}(0,1)$ are shown in Fig.\ 2.
No currents originating at $\p B(0,2)$ have access to the region $B(0,1)$, 
so that (heuristically) if the conductivity is changed in $B(0,1)$, the measurements on the boundary $\p B(0,2)$ do not change. Any object in $B(0,1)$ is both unaffected and undetectable by currents from the outside. {Moreover, 
all voltage-to-current measurements made on $\p B(0,2)$ give the same results as the measurements on
the surface of a ball filled with homogeneous, isotropic material.} The object is said to be \emph{cloaked}, and
the structure on $B(0,2)\setminus \overline{B}(0,1)$ producing this effect is said to be a \emph{cloaking device}.

However, this intuition needs to be supported by rigorous analysis of the solutions on the \emph{entire} region $B(0,2)$.
If we consider a singular metric $\widetilde g$ defined
by $(F_1)_*(g_0)$ on $B(0,2)\setminus \overline B(0,1)$, 
an arbitrary positive-definite symmetric metric on $B(0,1)$, and  $H^1(B(0,2))$ smooth solutions
of the conductivity equation, it was shown
in \cite{GLU2,GLU3} that, for $n\ge3$,  the following theorem holds.

\begin{theorem}\label{main}
$\Lambda_{\widetilde g}= \Lambda_{g_0}.$
\end{theorem}

In other words the boundary observations for the singular metric $\widetilde g$
are the same as the boundary observations for the Euclidean metric;  thus,
any object in $B(0,1)$ is invisible to electrostatic measurements.
We remark here that the measurements of the DN map or ``near field" are
 equivalent to scattering or ``far field" information \cite{Ber}. Also, see \cite{KSVW} for the planar case, $n=2$.

In the proof of Thm. \ref{main} one has to pay special attention  to
what is meant by a solution of the Laplace-Beltrami equation (\ref{eqn1.9})
with singular coefficients. In \cite{GLU2,GLU3}, we considered functions that
are  bounded and in the Sobolev space $H^1(B(0,2))$, and are solutions in the sense of distributions.  
Later, we will also consider  more general solutions.

The proof of Theorem \ref{main} has two ingredients, which are also the main
ideas behind transformation optics:

\begin{itemize}
\item The invariance  of the equation under transformations, \ie, identity (\ref{eqn1.11}). 

\item A (quite standard) removable 
singularities theorem: points are removable singularities of bounded harmonic functions.

\end{itemize}

The second point implies that bounded solutions of the Laplace-Beltrami equation with the singular metric indicated
 above on the annulus $B(0,2)\setminus B(0,1)$ are equivalent to bounded harmonic functions on the whole ball $B(0,2)$.
This shows that any $H^1$ solution $u$ to the equation (\ref{eqn1.8}) is constant on the ball of radius 1 with the
 constant the value of the corresponding harmonic function $v(0)$ with $v=u \circ F_{1}.$

The 2003 papers \cite{GLU2,GLU3} were intended to give counterexamples
to uniqueness in Calder\'on's problem when the anisotropic conductivity
is allowed to be only  positive semi-definite. 
 In the summer of 2006, Bob Kohn called our attention to the paper \cite{PSS1}
where the same transformation $F_1$ was used to propose cloaking for
Maxwell's equations, justified by the analogue of (\ref{eqn1.11}). In fact the electrical permittivity and magnetic permeability in the blueprint for a cloaking device given in \cite{PSS1} are

\begin{equation}
(\tilde \e^{ij})=(\tilde \mu^{ij})=(|\tilde g|^{1/2}\tilde g^{ij})
\end{equation}

with $\tilde g=(F_1)_{*} g_0$. The proposal of \cite{Le} (appearing in the same issue of \emph{Science}!) uses 
a different construction in two 
dimensions with explaining the behavior of the light rays but not the electromagnetic waves. The argument of \cite{PSS1}
is only valid outside the cloaked region; it doesn't take into account the behavior of the waves on the entire region, including the cloaked region and  its boundary, the cloaking surface.  In fact,
the sequel  \cite{CPSSP}, which
gave  numerical simulations of the electromagnetic waves in the presence of
a cloak, states: ``Whether perfect cloaking is
achievable, even in theory, is also an open question". In
\cite{GKLU1} we established that  perfect cloaking is indeed 
mathematically
 possible at any fixed frequency.

Before we discuss the paper \cite{GKLU1} and other developments,
we would like to point out that it is still
an open  question  whether visual cloaking is feasible in practice, \ie, whether
one can realize such theoretical blueprints for cloaking over all, or some large portion of, the visible spectrum. 
The main  experimental evidence has been at microwave frequencies
\cite{Sc}, with a limited version at a visible frequency \cite{smol}.
While significant progress has been made in the design and fabrication of metamaterials, including recently for visible light  \cite{Liu,Shv},
 metamaterials are nevertheless very dispersive  and one expects them to work only for a narrow range of frequencies. 
Even theoretically, one  can unfortunately not expect to actually
cloak electromagnetically  at all frequencies, since the group velocity cannot be faster than the velocity of light in a vacuum.

In  \cite{GKLU1},  Thm. \ref{main} was  extended to the Helmholtz equation, which models scalar optics (and acoustic waves \cite{Ch3,Cu} and quantum waves under
some conditions \cite{Zhang}), and Maxwell's equations, corresponding to  invisibility for general electromagnetic waves. The  case of acoustic or electromagnetic
sources inside and outside the cloaked region, leading to
serious obstacles to cloaking for Maxwell's equations, was also treated.

In Sec. \ref{subsec-Helmholtz}, we consider acoustic cloaking, \ie,
cloaking for the  Helmholtz equation at any non-zero frequency with an acoustic
source $\rho$,

\begin{equation}
(\Delta_g + k^2)u=\rho,\quad \mbox {in } B(0,2)\, .
\end{equation}
Physically, the anisotropic density is given by $|g|^{1/2} g^{ij}$  and the bulk modulus by $|g|^{1/2}.$

For acoustic cloaking, even with acoustic sources inside $B(0,1)$, we consider 
the same singular metric considered for electrostatics. However, we need to change the notion of a solution 
since for a generic frequency a $H^1 (B(0,2))$ smooth 
solution of the Helmholtz equation cannot 
simultaneously satisfy a homogeneous Neumann
condition on the surface of the cloaked region \cite[Thm.\ 3.5]{GKLU1}
and have a Dirichlet boundary value that is a non-zero constant.
We change the notion of solution for Helmholtz equation to a {\sl finite energy
solution} (see Sec. 4.3). The key ingredient of 
the rigorous justification of 
transformation optics is then  a removable singularities theorem for the Laplacian on $H^1(B(0,2)\setminus 0)$.  

In Sec.\ \ref{subsec-Maxwell} we consider the case of Maxwell's equations.
In the absence of internal currents, the construction of \cite{GLU2,GLU3}, called the \emph{single coating}  in \cite{GKLU1}, still works  once one makes an appropriate definition of {\sl
finite energy solutions}.
However,  cloaking using this construction fails in the presence of sources within the cloaked region, \ie,
for cloaking of active objects, 
due to the nonexistence of finite energy, distributional solutions.
This problem can 
be avoided by augmenting the external metamaterial layer with an
appropriately matched internal one in $D$;  this is called the \emph{double
coating}; see Sec.\  \ref{subsec-Maxwell}.

 In Sec.\ 5 we consider  another type of transformation optics-based device,  an
{\sl electromagnetic wormhole}. The idea is to create a secret connection
between two points in space so that only the incoming and the outgoing waves are visible. One tricks the electromagnetic waves to behave as though they were propagating on a handlebody, giving the impression that the topology of  space has been changed. Moreover, one can manipulate the rays travelling inside the handle
to obtain various additional 
optical effects;  see Fig. 3. Mathematically this is accomplished by using
the single coating construction with special boundary conditions on the cloaking surface.
The main difference is that,  instead of a point,  we blow up a curve, which in dimension $3$ or higher is also an $H^1$ removable singularity for solutions of Maxwell's equations.

\begin{figure}[htbp]\label{ray tracing 1}
\begin{center}
\centerline{\includegraphics[width=.375\linewidth]
{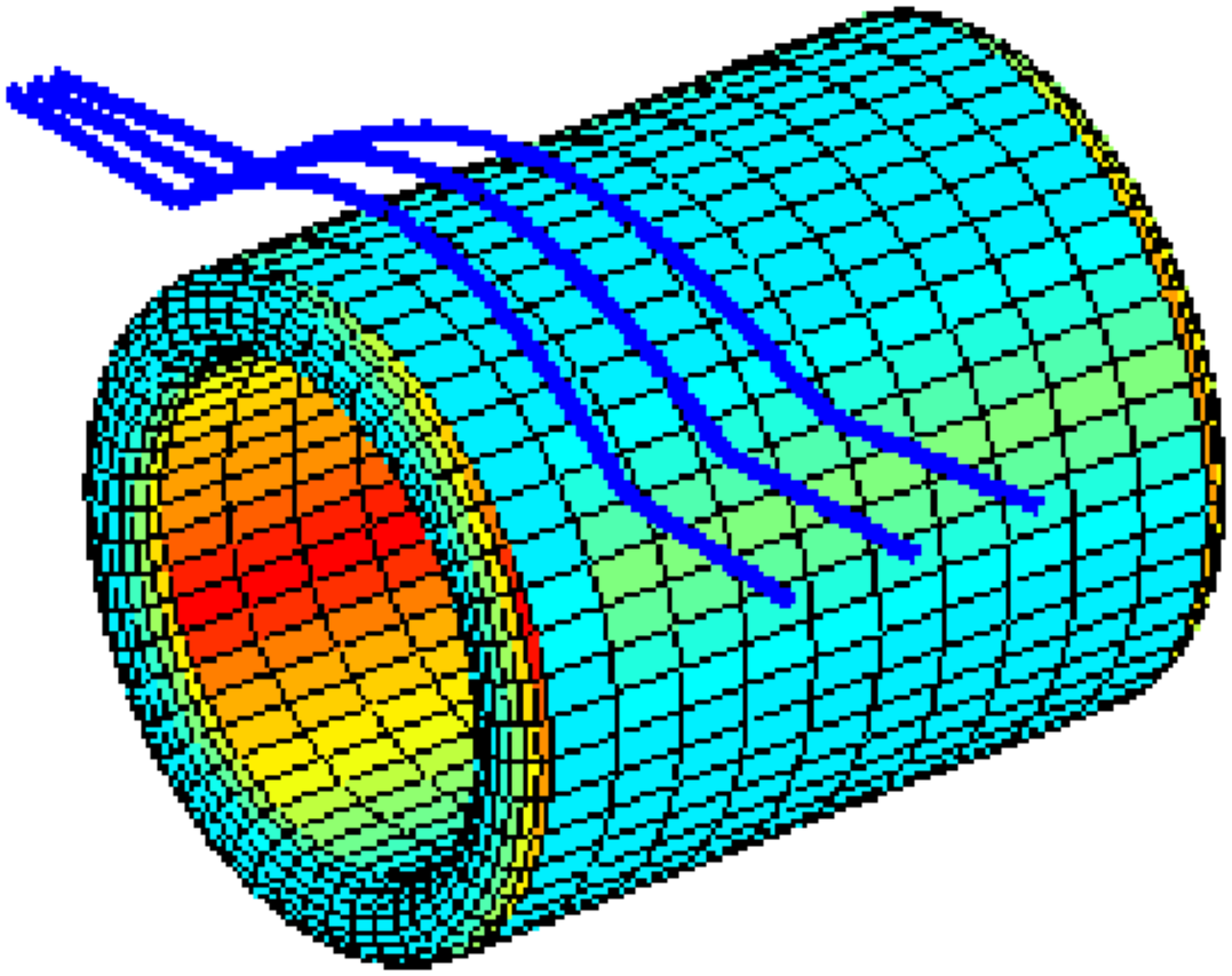}\hspace{10mm}
\includegraphics[width=.5\linewidth]{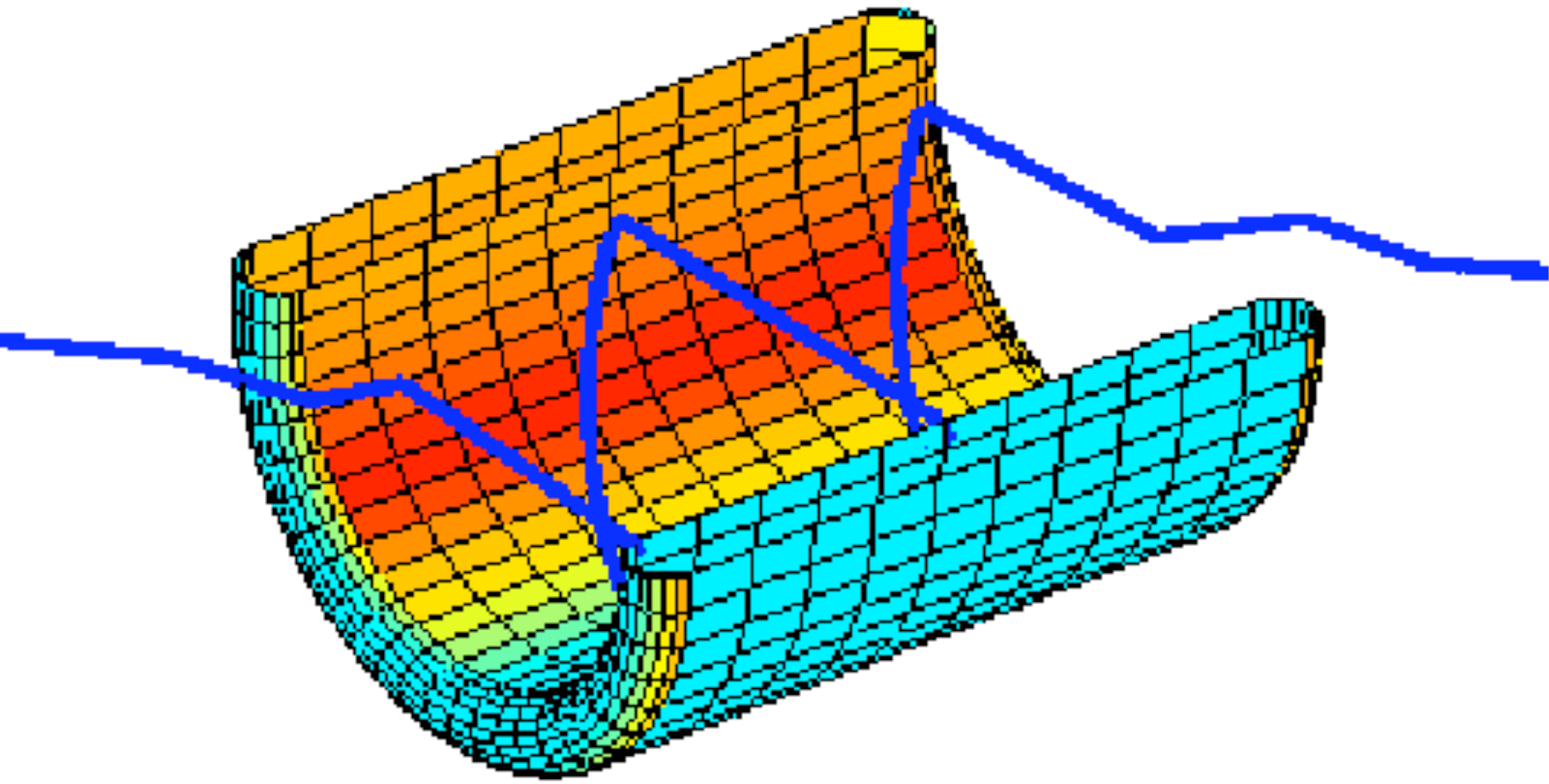}}
\end{center}
\caption{An electromagnetic wormhole is obtained by blowing up a metric near
a curve. This corresponds to $\epsilon$ and $\mu$ on the exterior of a thickened cylinder
causing electromagnetic  waves to propagate as if  a handle were attached
to  Euclidean space. Behavior of light rays:
{\bf (Left)}  Rays travelling outside wormhole. {\bf (Right)} A ray  transiting wormhole.}
\end{figure}

Both the anisotropy and singularity of the cloaking devices present
serious challenges in trying to physically realize such theoretical plans
using metamaterials. In Sec.\ 7, we give a  general method,
{\sl isotropic transformation optics}, for dealing with both of these
problems;
we describe it  in some detail in the context of cloaking, but it should be
applicable to a wider range of transformation optics-based designs.

A well known
phenomenon in effective medium theory is  that homogenization of isotropic
material parameters  may lead, in the small-scale limit, to anisotropic ones
\cite{Milt}. Using ideas from \cite{Allaire,Cherka} and elsewhere,  we showed
in \cite{GKLU6,GKLU7,GKLU8} how to
exploit this to find
cloaking material parameters that are at once both  isotropic and
nonsingular, at the
price of replacing perfect cloaking with  {\it approximate} cloaking of
arbitrary accuracy.
This  method, starting with  transformation optics-based
designs and constructing  approximations to them, first by {\it
nonsingular}, but still anisotropic, material parameters, and
then by nonsingular {\it  isotropic}  parameters,  seems to be a very
flexible tool for
creating physically realistic designs, easier to  implement than the ideal
ones due to the relatively tame nature of the materials needed,   yet
essentially capturing the desired effect on waves for all practical purposes.

In Sec. 8 we consider some  further developments and open problems.

\section{Visibility for electrostatics:  \\ Calder\'on's
problem}\label{sec-eit}

Calder\'on's inverse conductivity problem, which forms the mathematical
foundation of
Electrical Impedance Tomography (EIT),  is the
question of whether an  unknown conductivity distribution inside a domain
in $\R^n$, modelling, \eg,
the Earth, a human thorax, or a manufactured part, can be
determined  from voltage and current
measurements made on the boundary.
A.P. Calder\'on's motivation for proposing this problem  was
geophysical prospection. In
the 1940's, before his distinguished career as a mathematician, Calder\'on
was an engineer working for the Argentinian state oil company.
Apparently,  Calder\'on had already at that time formulated
the problem that now bears his name, but he did
not publicize this work until thirty years later \cite{C}.

One widely studied
potential application of EIT is
the early diagnosis of breast cancer \cite{CIN}.  The conductivity of a
malignant breast tumor is typically 0.2 mho,  significantly
higher than normal tissue, which has been typically measured at 0.03 mho.
See the book \cite{Ho} and the special issue of \emph{Physiological
Measurement}
\cite{HIMS} for applications of EIT to medical imaging and other fields, and
\cite{B} for a review.

\vspace{.75cm}

\begin{figure}[htbp]\label{EIT figure}
\begin{center}
\centerline{\includegraphics[width=.4\linewidth]{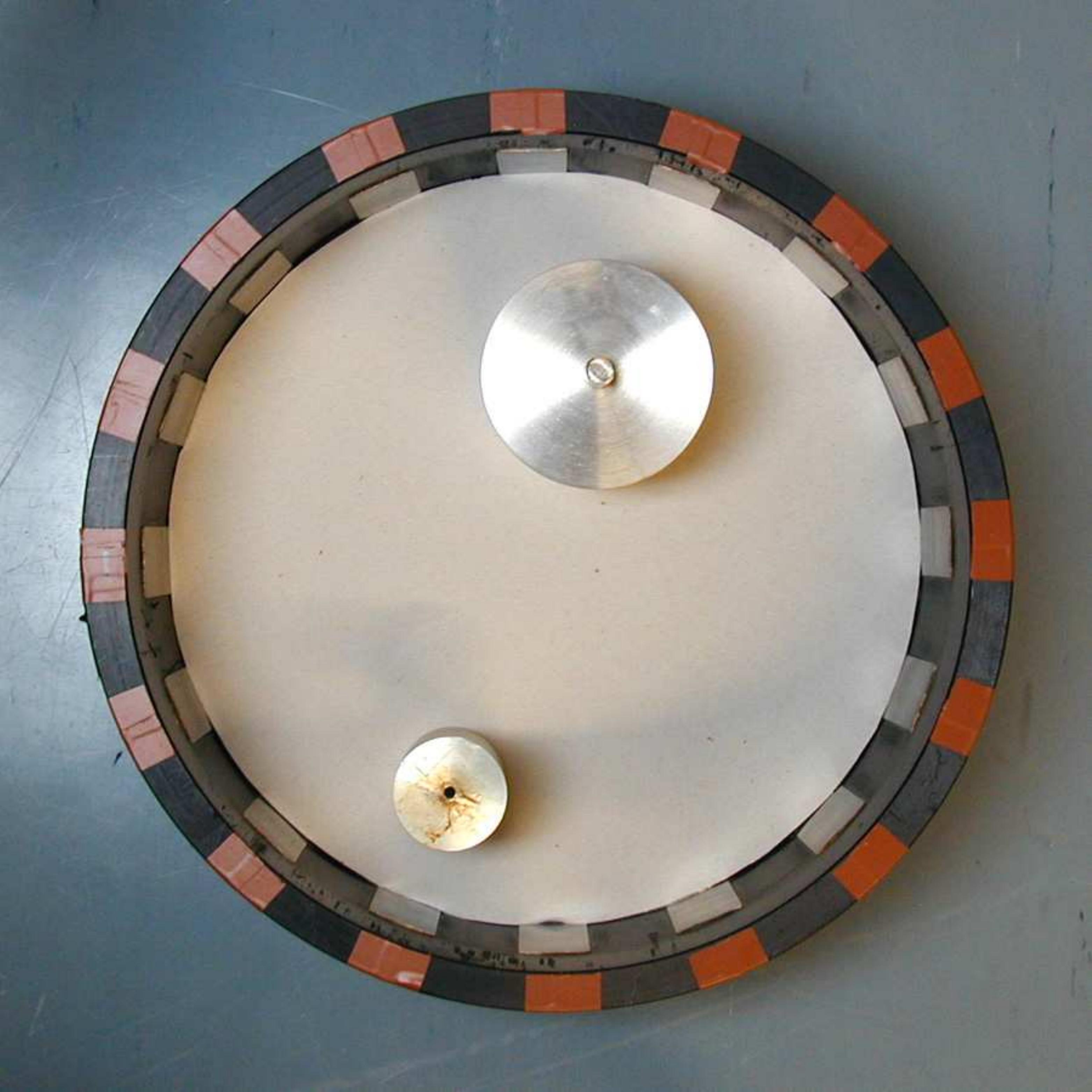}\hspace{1.0cm}
\includegraphics[width=.4\linewidth]{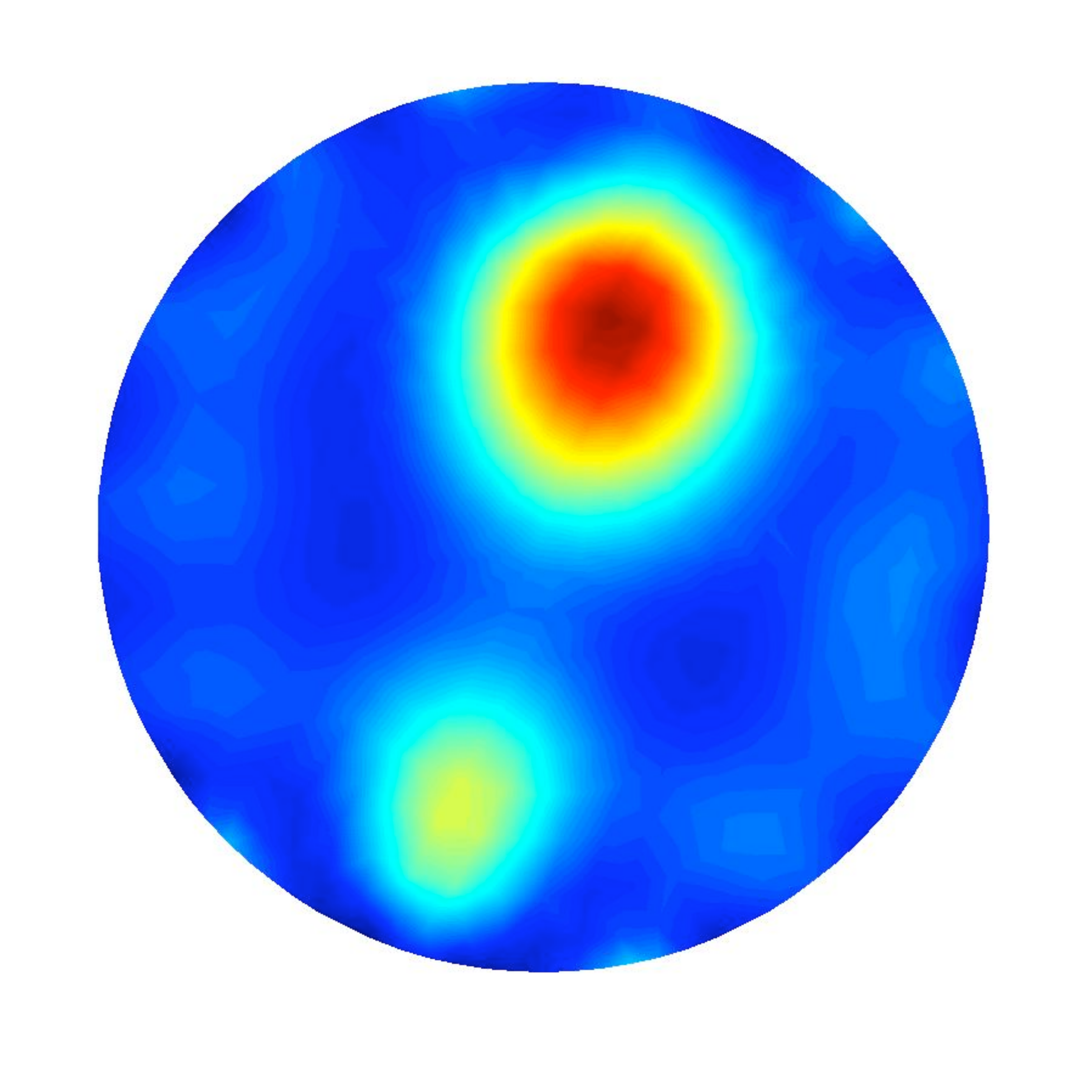}}
\end{center}
\caption{{\bf Left:} An EIT measurement configuration for imaging
objects in a tank.  The electrodes used for measurements
are at the boundary of the tank, which is filled with a conductive liquid.
{\bf Right:}
A reconstruction of the conductivity inside the tank obtained
using boundary measurements. \qquad[Jari Kaipio, Univ. of Kuopio, Finland;
by permission.]}
\end{figure}

For isotropic conductivities this problem can be mathematically
formulated as follows: Let $\Omega$ be the measurement domain, and
denote by $\sigma(x)$  the coefficient, bounded  from above and below by
positive constants, describing
the electrical conductivity in $\Omega$.
In $\Omega$ the voltage potential $u$ satisfies  a
divergence form equation,
\begin{equation}\label{johty}
\nabla\cdot \sigma\nabla u = 0.
\end{equation}

To uniquely fix the solution $u$ it is enough to give its value, $f$, on the
boundary.  {In the idealized case, one measures,  for
all voltage distributions $u|_{\p \Omega}=f$ on the boundary the
corresponding
current fluxes, $\nu\cdotp \sigma \nabla u$, over the entire boundary,
  where $\nu$ is the exterior unit normal to
$\p\Omega$.} Mathematically this amounts to the knowledge of the
Dirichlet-to-Neumann (DN) map,  $\Lambda_\sigma$,
corresponding to $\sigma$, \ie, the map taking the Dirichlet boundary
values of the solution to (\ref{johty}) to the
corresponding Neumann boundary values,

\begin{equation}
\Lambda_\sigma: \ \ u|_{\p \Omega}\mapsto \nu\cdotp \sigma \nabla u|_{\p
\Omega}.
\end{equation}
{Calder\'on's inverse problem is then to reconstruct $\sigma$ from
$\Lambda_\sigma$.

In the following subsections, we give a brief overview of the positive
results known for Calder\'on's problem and  related inverse problems. 

{A basic distinction, important for understanding cloaking, is
between \emph{isotropic} conductivities, which are scalar-valued, and
\emph{anisotropic} conductivities, which are symmetric matrix- or
tensor-valued, modelling situations where the conductivity depends on both
position and direction. Of course, an isotropic $\sigma(x)$ can be
considered as anisotropic by identifying it with $\sigma(x) I_{n\times n}$.}

Unique determination of an isotropic conductivity from the DN map
was shown in dimension $n>2$ for $C^2$ conductivities in \cite{SU}.
At the writing of the current paper this result has been extended to
conductivities having $\frac{3}{2}$ derivatives in \cite{BT} and
\cite{PPU}. In two dimensions the first unique identifiability
result was proven in \cite{N1} for $C^2$ conductivities. This was
improved to Lipschitz conductivities in \cite{BU} and to merely
$L^\infty$ conductivities in \cite{AP}. All of these results use
{\sl complex geometrical optics} (CGO) solutions, 
the construction of which we review in Sec. \ref{sec cgo}.
{We briefly discuss in Sec. \ref{sec quantum shield}
\emph{shielding}, a less satisfactory variant of cloaking which is possible
using highly singular isotropic materials.}

In Sec. \ref{sec aniso cond} we discuss the case of anisotropic
conductivities,
\ie, conductivities that may vary not only with location 
but also on the direction. In this case,
the problem is invariant under changes of variables that are the
identity at the boundary. We review  the positive results that are known
about the Calder\'on problem in this setting. The fact that the anisotropic
conductivity
equation is invariant under transformations plays a crucial role on
the constructions of electromagnetic parameters that make objects
invisible, but for those one needs to make a final leap to using
\emph{singular} transformations.

\subsection{Complex geometrical optics solutions}\label{sec cgo}

In this section, we consider isotropic conductivities.
If $u$ is a solution of (\ref{johty}) with boundary data $f$, the
divergence theorem gives that
\begin{equation}\label{eq:0.5}
Q_\sigma (f):=\int_\Omega \sigma|\nabla u|^2 \,
dx=\int_{\partial\Omega}\Lambda_\sigma(f)f\,dS
\end{equation}
where $dS$ denotes surface measure. In other words $Q_\sigma(f)$ is
the quadratic form associated to the linear map $\Lambda_\sigma(f)$,
i.e., to know $\Lambda_\sigma(f)$ or $Q_\sigma(f)$ for all $f\in
H^{\frac{1}{2}} (\partial\Omega)$ is equivalent. The form $Q_\sigma
(f)$ measures the energy needed to maintain the potential $f$ at the
boundary. Calder\'on's point of view
in order to determine $\sigma$ in $\Omega$
was to find enough solutions
$u\in H^1(\Omega)$ of the conductivity equation
$\mbox{div}(\sigma\nabla u)=0$ so that the  functions $|\nabla u|^2$ span
a dense set  (in an appropriate topology). Notice
that the DN map (or $Q_\sigma$) depends non-linearly on $\sigma.$
Calder\'on considered the linearized problem at a constant
conductivity.  A crucial ingredient in his approach is the use of
the harmonic complex exponential solutions:

\begin{equation}\label{eq:1.3}
u=e^{x\cdot\rho}, \mbox{ where }\rho\in\C^n \mbox{ with }
\rho\cdot\rho=0.
\end{equation}

Sylvester and Uhlmann \cite{SU} constructed in dimension $n\ge
2$ complex geometrical optics (CGO) solutions of the conductivity
equation for $C^2$ conductivities similar to Calder\'on's.  This can
be reduced to constructing solutions in the whole space (by
extending $\sigma=1$ outside a large ball containing $\Omega$) for
the Schr\"odinger equation with potential. We describe this more
precisely below.

Let $\sigma\in C^2(\R^n)$, $\sigma$ strictly positive in $\R^n $
and $\sigma=1$ for $\vert x \vert \ge R$, for some $R>0$. Let $L_\sigma
u=\nabla\cdot\sigma\nabla u$. Then we have
\begin{equation}\label{eq:2.1}
\sigma^{-\frac{1}{2}}L_\sigma \left(\sigma^{-\frac{1}{2}} v\right)=
(\Delta-q)v,
\end{equation}
where 
\begin{equation}\label{eq:2.2}
q=\frac{\Delta\sqrt{\sigma}}{\sqrt{\sigma}}.
\end{equation}
Therefore,  to construct solutions of $L_\sigma u=0$ in $\R^n$ it
is enough to construct solutions of the Schr\"odinger equation
$(\Delta-q) v=0$ with $q$ of the form (\ref{eq:2.2}). The next result
proven in \cite{SU}  states the existence of complex
geometrical optics solutions for the Schr\"odinger equation
associated to any bounded and compactly supported potential.

\begin{theorem}\label{thm:2.1}
Let $q\in L^\infty(\R^n)$, $n\ge 2$, with $q(x)=0$ for $|x|\ge
R>0$. Let $-1<\delta<0.$  There exists $\epsilon(\delta)$ and such
that for every $\rho \in \C^n$ satisfying
$$\rho \cdot \rho =0$$ and
$$ {\Vert (1+ \vert x \vert^2)^{1/2} q\Vert_{L^\infty (\R^n)} +1 \over \vert
\rho \vert} \le \epsilon$$
there exists a unique solution to
$$(\Delta-q)v=0$$ of the form
\begin{equation}\label{eq:2.4}
v=e^{x\cdot\rho}(1+\psi_q(x, \rho))
\end{equation}
with $\psi_q(\cdot, \rho)\in L^2_\delta(\R^n)$. Moreover
$\psi_q(\cdot, \rho)\in H^2_\delta(\R^n)$ and, for $0\le s\le 1$,
there exists $C=C(n, s,\delta)>0$ such that
\begin{equation}\label{eq:2.5}
\Vert\psi_q(\cdot, \rho)\Vert_{H^s_\delta}\le \frac{C}{|\rho|^{1-s}}
\end{equation}
\end{theorem}

Here $$L^2_\delta(\R^n)=\{f; \int(1+|x|^2)^\delta|f(x)|^2dx <\infty \}$$
with the norm given by $\Vert
f\Vert^2_{L^2_\delta}=\int(1+|x|^2)^{\delta}|f(x)|^2dx$ and
$H^m_\delta(\R^n)$ denotes the corresponding Sobolev space. Note
that for large $|\rho|$ these solutions behave like Calder\'on's
exponential solutions $e^{x \cdot\rho}$. The equation for $\psi_q$
is given by
\begin{equation}\label{eq:2.6}
(\Delta+2\rho\cdot\nabla)\psi_q=q(1+\psi_q).
\end{equation}
The equation (\ref{eq:2.6}) is solved by constructing an inverse for
$(\Delta+2\rho\cdot\nabla)$ and solving the integral equation
\begin{equation}\label{eq:2.7}
\psi_q=(\Delta+2\rho\cdot \nabla)^{-1}(q(1+\psi_q)).
\end{equation}

\begin{lemma}\label{lem:2.2}
Let $-1<\delta<0, \quad 0\le s\le 1$. Let $\rho\in\C^n\setminus 0$,
$\rho\cdot\rho=0$. Let $f\in L^2_{\delta+1}(\R^n)$. Then there
exists a unique solution $u_\rho\in L^2_\delta(\R^n)$ of the
equation
\begin{equation}\label{eq:2.8}
\Delta_\rho u_\rho:=(\Delta+2\rho\cdot\nabla)u_\rho=f.
\end{equation}

Moreover $u_\rho\in H^2_\delta(\R^n)$ and
\[
\Vert u_\rho\Vert_{H^s_\delta(\R^n)} \le \frac{C_{s ,\delta}\Vert
f\Vert_{L^2_{\delta+1}}}{|\rho|^{1-s}}
\]
for $0\le s\le 2$ and for some constant $C_{s,\delta}>0$.
\end{lemma}

The integral equation (\ref{eq:2.7}) with 
Faddeev's Green kernel \cite{faddeev}
 can then be solved in
$L^2_\delta(\R^n)$ for large $|\rho|$ since
\[
(I-(\Delta+2\rho\cdot\nabla)^{-1}q)\psi_q=(\Delta+2\rho\cdot\nabla)^{-1}q
\]
and $\Vert(\Delta+2\rho\cdot\nabla)^{-1}q\Vert_{L^2_\delta\to
L^2_\delta}\le \frac{C}{|\rho|}$ for some $C>0$ where
$\Vert\cdot\Vert_{L^2_\delta\to L^2_\delta}$ denotes the operator
norm between $L^2_\delta(\R^n)$ and $L^2_\delta(\R^n)$. We will
not give details of the proof of Lemma \ref{lem:2.2} here. We refer
to the papers \cite{SU,SU86} .

If $0$ is not a Dirichlet eigenvalue for the Schr\"odinger equation
we can also define the DN map
$$\Lambda_q(f)= \frac{\partial u}{\partial \nu}|_{\partial\Omega}$$
where $u$ solves
$$(\Delta-q)u=0; \quad u|_{\partial\Omega}=f.$$
Under some regularity assumptions,
the DN map associated to the Schr\"odinger equation $\Delta -q$
determines in dimension $n>2$ uniquely a bounded potential; see
\cite{SU} for the smooth case, \cite{NSU} for $L^\infty$, and
\cite{Chanillo} for  potentials in a Fefferman-Phong class.

The two dimensional results of \cite{N1},\cite{BU}, \cite{AP} use
similar CGO solutions and the $\overline \partial$ method in the
complex frequency domain, introduced by Beals and Coifman in \cite{BC1} and
generalized to higher dimensions in several articles
\cite{BC22},\cite{ABF},\cite{NA}.

More general CGO solutions have been constructed in \cite{KSU}
of the form
\begin{equation}\label{cgo}
u= e^{\tau(\phi+i\psi)} (a+r),
\end{equation}
where $\nabla \phi \cdot\nabla \psi=0, |\nabla \phi|^2=|\nabla
\psi|^2$ and $\phi$ is a limiting Carleman weight (LCW). Moreover $a$ is
smooth and
non-vanishing and $\Vert r\Vert_{L^2(\Omega)} =O(\frac{1}{\tau})$,
$\Vert r\Vert_{H^1(\Omega)} =O(1)$. Examples of LCW are the linear
phase $\phi(x)=x\cdot\omega, \omega\in S^{n-1},$ used in the results
mentioned above,
and the non-linear phase $\phi(x)= \ln|x-x_0|$, where  $x_0\in {\bf
R}^n\setminus\overline{{\rm ch\,}(\Omega )}$  ($ch(\cdot)$ denoting the
convex hull) which was used in
\cite{KSU} for the problem where the DN map is measured in parts of the
boundary.
For a characterization of all the LCW in $\R^n$, $n>2$, see
\cite{DKSU}. In two dimensions any harmonic function is a LCW
\cite{UW}.

Recently, Bukhgeim \cite{Buk} used  CGO solutions in two dimensions of the
form
(\ref{cgo})
with $\phi= z^2$ or  $\phi=\overline z^2$ (identifying $\R^2\sim\mathbb C$)
to prove
that any compactly supported potential $q\in L^p, p>2$, is uniquely
determined by  Cauchy data of the associated Schr\"odinger operator.

Other applications to inverse problems using the CGO solutions
described above with a linear phase are:

\begin{itemize}
\item {\bf Quantum scattering:} It is shown in  \cite{N2} and \cite{No}
that in dimension $n>2$ the scattering amplitude at a fixed energy
determines uniquely a two body compactly supported potential. This
result also follows from \cite{SU} (see for instance \cite{U2},
\cite{U3}). Applications of CGO solutions to the 3-body problem were
given in \cite{VU}. In two dimensions the result of \cite{Buk} implies
unique determination of the potential from the scattering amplitude at
fixed energy.

\item {\bf Scalar optics:} The DN map associated to the Helmholtz equation
$\Delta+k^2n^2(x)$ with an isotropic index of refraction $n$ determines
uniquely
a bounded index of refraction in dimension $3$ or larger, see e.g.\ 
\cite{N2,SU}.

\item {\bf Optical tomography in the diffusion approximation:} In this case
we have $\nabla \cdot a(x)\nabla u- \sigma_a(x)u -i \omega u=0$ in
$\Omega$ where $u$ represents the density of photons, $a(x)$ the
diffusion coefficient, and $\sigma_a$ the optical absorption. Using
the result of \cite{SU} one can show in dimension three or higher that if
$\omega\neq 0$ one can
recover both $a$ and $\sigma_a$ from the corresponding DN map. If
$\omega=0$ then one can recover one of the two parameters.

\item {\bf Electromagnetics:} The DN map for isotropic Maxwell's equations
determines uniquely the isotropic
electric permittivity, magnetic permeability and conductivity
\cite{OPS}. This system can in fact be reduced to an $8\times 8$
Schr\"odinger
system, $\Delta\cdot I_{8\times 8} -Q$  \cite{OPS}.
\end{itemize}

For further discussion and other applications of CGO with linear
phase solutions, including inverse problems for the magnetic Schr\"odinger
operator,  see \cite{U2}.

\subsection{Quantum Shielding}\label{sec quantum shield}

In \cite{GLU1}, also using  CGO solutions, we proved uniqueness for the
Calder\'on problem for  Schr\"odinger operators having  a more singular
class of potentials, namely potentials conormal 
to  submanifolds of $\R^n, n\ge 3$. These may be more singular than the
potentials in 
\cite{Chanillo} and, for the case of a hypersurface $S$, can have any
strength less than the delta function $\delta_S$.

However, for much more singular potentials, there are counterexamples to
uniqueness. We constructed a class of potentials that shield any information
about the
 values of a potential
on a region $D$ contained in a domain $\Omega$ from measurements of
solutions at $\p\Omega$. In other words, the boundary information
obtained outside the shielded region is independent of $q|_D$.
On $\Omega\setminus D$, these potentials behave like $q(x)\sim -C
d(x, \partial D)^{-2-\epsilon}$ where $d$ denotes the distance to $\partial
D$
and $C$ is a positive constant. In $D$, Schr\"odinger's cat could live
forever.
From the point of view of quantum  mechanics, $q$ represents a potential
barrier so steep that no tunneling can occur. From the point of view of
optics and acoustics, no sound waves or electromagnetic waves will
penetrate, or emanate from, $D$. However, this construction should be
thought of as shielding, not cloaking, since the potential barrier that
shields  $q|_D$ from boundary observation
is itself detectable.

\subsection{Anisotropic conductivities}\label{sec aniso cond}

Anisotropic conductivities depend on direction. Muscle tissue in the
human body is an important example of an anisotropic conductor. For
instance cardiac muscle has a conductivity of 2.3 mho in the
transverse direction and 6.3 in the longitudinal direction.  The
conductivity in this case is represented by a positive definite,
smooth, symmetric matrix $\sigma=(\sigma^{ij}(x))$ on $\Omega$.

Under the assumption of no sources or sinks of current in $\Omega$,
the potential $u$ in $\Omega$, given a voltage potential $f$ on
$\partial\Omega$, solves the Dirichlet problem
\begin{equation}\label{eqn1.1}
\left\{\begin{array}{rcl}
\nabla \cdot \sigma \nabla u :=\mathop{\sum}\limits^{n}_{i,j=1}\frac{\p}{\p
x_i}\left(\sigma^{ij}\frac{\p u}{\p
x_j}\right) & = & 0\hbox{ on }\Omega \\
u|_{\partial\Omega} & = & f.
\end{array}\right.
\end{equation}
The DN map is defined by
\begin{equation}\label{eqn1.2}
\Lambda_\sigma(f)=\sum^n_{i,j=1}\nu^i\sigma^{ij}\frac{\p u}{\p
x_j}\Big|_{\partial\Omega}
\end{equation}
where $\nu=(\nu^1, \ldots, \nu^n)$ denotes the unit outer normal to
$\partial\Omega$ and $u$ is the solution of (\ref{eqn1.1}).  The
inverse problem is whether one can determine $\sigma$ by knowing
$\Lambda_\sigma$.  Unfortunately, $\Lambda_\sigma$ doesn't determine
$\sigma$ uniquely. This observation is due to L.\ Tartar (see
\cite{KV2} for an account).

Indeed, let $\psi:\overline\Omega\to\overline\Omega$ be a $C^\infty$
diffeomorphism with $\psi|_{\partial\Omega}= Id$, the identity map.  We have

\begin{equation}
\Lambda_{\widetilde\sigma}=\Lambda_{\sigma}
\end{equation}
where {$\widetilde\sigma=\psi_*\sigma$ is the push-forward
of conductivity $\sigma$ in $\psi$, }

\begin{equation}\label{eqn1.6}
\psi_*\sigma=\left (\frac{(D\psi)^T\circ\sigma\circ(D\psi)}
{|\hbox{det} D\psi|}\right )\circ\psi^{-1}.
\end{equation}

Here $D\psi$ denotes the (matrix) differential of $\psi$,
$(D\psi)^T$ its transpose and the composition in (\ref{eqn1.6}) is
to be interpreted as multiplication of matrices.

We have then a large number of conductivities with the same DN map:
any change of variables of $\Omega$ that leaves the boundary fixed
gives rise to a new conductivity with the same electrostatic
boundary measurements.

 The question is then whether this is the only
obstruction to unique identifiability of the conductivity. In two
dimensions, this was
proved for $C^3$ conductivities by reducing the anisotropic problem to
the isotropic one by using isothermal coordinates \cite{S} and using
Nachman's isotropic result \cite{N1}. The regularity was improved in
\cite{SuU}
to Lipschitz conductivities using the techniques of \cite{BU} and to
$L^\infty$ conductivities in \cite{ALP} using the results of
\cite{AP}.

In the case of dimension $n\ge 3$, as was pointed out in \cite{LeU},
this is a problem of geometrical nature and makes sense for general
compact Riemannian manifolds with boundary.

{Let $(M,g)$ be a compact Riemannian manifold with boundary; the
Laplace-Beltrami operator associated to the metric $g$ is given in
local coordinates by
(\ref{eqn1.8}).
Considering the Dirichlet problem (\ref{eqn1.9}) associated to
(\ref{eqn1.8}),
we defined in the introduction the DN map in this case by}
\beq\label{eqn1.10 B}
\Lambda_g (f)=\sum^n_{i,j=1}\left.\nu_i g^{ij} \frac {\partial u}
{\partial x_j}
\sqrt{|g|}\right|_{\partial\Omega}
\eeq
where $\nu$ is the unit-outer normal.

The inverse problem is to recover $g$ from $\Lambda_g$.

If $\psi$ is a $C^\infty$ diffeomorphism of $\overline M$ which
is the identity on the boundary, and $\psi^\ast g$ denotes the
pull back of the metric $g$ by  $\psi$, we then have that
(\ref{eqn1.11}) holds.

In the case that $M$ is an open, bounded subset of $\R^n$ with
smooth boundary, it is easy to see (\cite{LeU}) that for $n\ge 3$
\begin{equation}\label{eqn1.12}
\Lambda_g=\Lambda_\sigma,
\end{equation}
where
\begin{equation}\label{eqn1.13}
g^{ij}=|\sigma|^{-1/(n-2)}\sigma^{ij}, \quad
\sigma^{ij}=|g|^{\frac{1}{2}}g^{ij}.
\end{equation}

In the two dimensional case there is an additional obstruction since
the Laplace-Beltrami operator is conformally invariant. More
precisely we have
\[
\Delta_{\a g}=\frac{1}{\a}\Delta_g
\]
for any function $\a$, $\a > 0$. Therefore we have that (for $n=2$ only)
\begin{equation}\label{eqn1.14}
\Lambda_{\alpha(\psi^\ast g)}=\Lambda_{g}
\end{equation}
for any smooth function $\a > 0$ so that $\a|_{\p M}=1$.

Lassas and Uhlmann \cite{LU} proved that (\ref{eqn1.11}) is the
only obstruction to unique identifiability of the conductivity for
real-analytic manifolds in dimension $n\ge 3$. In the two
dimensional case they showed that (\ref{eqn1.14}) is the only
obstruction to unique identifiability {for $C^\infty$-smooth}
Riemannian
surfaces.  Moreover these results assume that $\Lambda$ is measured
only on an open subset of the boundary. We state the two basic
results.

Let $\Gamma$ be an open subset of $\p M$. We define for $f$, $\supp
f\subseteq\Gamma$
\[
\Lambda_{g,\Gamma}(f)=\Lambda_g(f)|_{\Gamma}.
\]

\begin{theorem}[$n\ge 3$]\label{thm2.2}
Let $(M,g)$ be a real-analytic compact, connected Riemannian
manifold with boundary. Let $\Gamma\subseteq\p M$ be real-analytic
and assume that $g$ is real-analytic up to $\Gamma$. Then
$(\Lambda_{g,\Gamma}, \partial M)$ determines uniquely $(M,g)$.
\end{theorem}

\begin{theorem}[$n=2$]\label{thm2.3}
Let $(M,g)$ be a compact Riemannian surface with boundary. Let
$\Gamma\subseteq\partial M$ be an open subset. Then
$(\Lambda_{g,\Gamma}, \partial M)$ determines uniquely the conformal
class of $(M,g)$.
\end{theorem}

Notice that these two results don't assume any condition on the
topology of the manifold except for connectedness. An earlier result
of \cite{LeU} assumed that $(M,g)$ was strongly convex and simply
connected and $\Gamma=\partial M$.
Theorem~\ref{thm2.2} was extended in \cite{LTU} to non-compact,
connected real-analytic manifolds with boundary.
The number of needed measurements for determination of the
conformal class for generic Riemannian surfaces was reduced
in \cite{HeM}. It was recently shown that Einstein manifolds are uniquely
determined up to isometry by the DN map \cite{GS}.

In two dimensions the invariant form of the conductivity equation is
given by

\begin{equation}
\mbox{div}_g(\beta \nabla_g)u:=g^{-1/2}\p_i \left(g^{1/2}\beta g^{ij}  \p_j u
\right)=0
\end{equation}
where $\beta$ is the conductivity and $\mbox{div}_g$ (resp.
$\nabla_g$) denotes divergence (resp. gradient) with respect to the
Riemannian metric $g.$ This includes the isotropic case considered by
Calder\'on with $g$ the Euclidian metric, and the anisotropic case by
taking $(g^{ij}=\gamma^{ij}$ and $\beta=|g|^{1/2}).$ It
was shown in \cite{SuU} for bounded domains of Euclidian space that
the isometry class of $(\beta,g)$ is determined uniquely by the
corresponding DN
map.

We remark that there is an extensive literature on a related inverse
problem, the
so-called Gelfand's problem,  where one studies the inverse
problem of determining a Riemannian manifold from the DN map associated
to the
Laplace-Beltrami operator for all frequencies, see \cite{KKL} and the
references cited there.

\section{Invisibility  
for Electrostatics}\label{sec-to for es}

The fact that  the boundary
measurements do not change, when a conductivity
is  pushed forward by
a smooth diffeomorphism  leaving the boundary fixed,
can already be considered
as a weak form of invisibility. Different media appear to be the same, and
the apparent location of objects  can change. However, this does not yet
constitute
real invisibility, as nothing has been hidden from view.

In  invisibility cloaking the aim is to
hide an object 
inside a domain by surrounding it with  (exotic) material
so that even the presence of this object can not be detected by 
measurements on the domain's boundary. This means that 
 all boundary 
measurements for the domain with this cloaked object included would be 
the same as if the domain were filled with a homogeneous, isotropic
material.
Theoretical models for this have been found by applying 
diffeomorphisms having singularities. 
These were first introduced in the framework of electrostatics, 
yielding counterexamples to the anisotropic Calder\'on problem
in the form of 
singular, anisotropic
conductivities in $\R^n, n\ge 3$,
indistinguishable
from a constant isotropic conductivity in that they have the same
Dirichlet-to-Neumann map  \cite{GLU2, GLU3}. The same construction was
rediscovered
for electromagnetism in \cite{PSS1}, with the intention of actually building
such a device with appropriately designed metamaterials; a modified version
of this was then experimentally demonstrated in \cite{Sc}. (See also
\cite{Le} for a somewhat different approach to cloaking in the high
frequency limit.)

The first constructions in this direction were based on  blowing
up the metric around a point \cite{LTU}. In this construction,
let $(M,g)$ be a compact 2-dimensional manifold with
non-empty boundary, let  
$x_0\in M$ and consider the manifold
\ba
\tilde M=M\setminus \{x_0\}
\ea
with the metric
\ba
\tilde g_{ij}(x)=\frac 1{d_M(x,x_0)^2}g_{ij}(x),
\ea
where $d_M(x,x_0)$ is the distance between $x$ and $x_0$ on $(M,g)$. 
Then $(\tilde M,\tilde g)$ is a complete, non-compact 2-dimensional 
Riemannian
manifold with the boundary $\p \tilde M=\p M$. 
Essentially, the point $x_0$ has been `pulled to infinity''.
On the manifolds $M$ and $\tilde M$ we consider the boundary value
problems
\ba
\left\{\begin{array}{l}
\Delta_g u=0\quad \hbox{in $M$,}\\
u=f\quad \hbox{on $\p M$,}\end{array}\right. \quad\hbox{and}\quad
\left\{\begin{array}{l}
\Delta_{\tilde g} \tilde u=0\quad \hbox{in $\tilde M$,}\\
\tilde u=f\quad \hbox{on $\p \tilde M$,}\\
\tilde u\in L^\infty(\tilde M).\end{array}\right.
\ea
These boundary value problems are uniquely solvable and
define the DN maps
\ba
\Lambda_{M,g}f=\p_\nu u|_{\p M},\quad
\Lambda_{\tilde M,\tilde g}f=\p_\nu \tilde u|_{\p \tilde M}
\ea
where $\p_\nu$ denotes the corresponding conormal derivatives.
Since, in the two dimensional case 
functions which are harmonic with respect to the metric $g$  stay harmonic
with respect to any
 metric which is conformal to $g$,  
one can see that
$\Lambda_{M,g}=\Lambda_{\tilde M,\tilde g}$. 
This can be seen using e.g.\ Brownian motion or 
capacity arguments. 
Thus, the boundary measurements for $(M,g)$ and $(\tilde M,\tilde g)$
coincide. 
This gives a counter example for the inverse
 electrostatic problem on Riemannian surfaces - even the topology
of possibly non-compact Riemannian surfaces can not be determined
using boundary measurements (see Fig.\ 5).

\begin{figure}[htbp]\label{LTU figure}
\begin{center}
\psfrag{1}{}
\psfrag{2}{}
\psfrag{3}{}
\includegraphics[width=12cm]{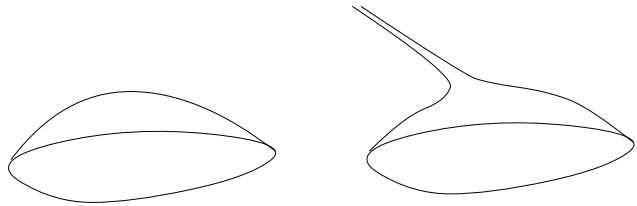} 
\vspace{-4cm}
\caption{Blowing up a metric at a point, after \cite{LTU}. The electrostatic
boundary measurements on the boundary of the  surfaces,
one compact and the other noncompact but complete, coincide.}
\end{center}
\end{figure}

The above example can be thought as a ``hole'' in a
Riemann surface that does not change the boundary measurements.
Roughly speaking, mapping the manifold $\tilde M$ smoothly to
the set $M\setminus \overline B_M(x_0,\rho)$, where $B_M(x_0,\rho)$
is a metric ball of $M$, and by putting an object in the 
obtained hole $\overline B_M(x_0,\rho)$, one could hide it from detection at the boundary.
This {observation was used} in \cite{GLU2,GLU3}, where ``undetectability"
results were  introduced in three dimensions, using
degenerations of Riemannian metrics, whose singular limits can be
considered as coming directly
from singular changes of variables. Thus, this construction can be
considered as an extreme, or singular, version of the transformation optics
of \cite{ward}.

\begin{figure}[htbp]\label{collapce}
\begin{center}
\psfrag{1}{}
\psfrag{2}{}
\psfrag{3}{}
\hspace{6cm}\includegraphics[width=1.0\linewidth]{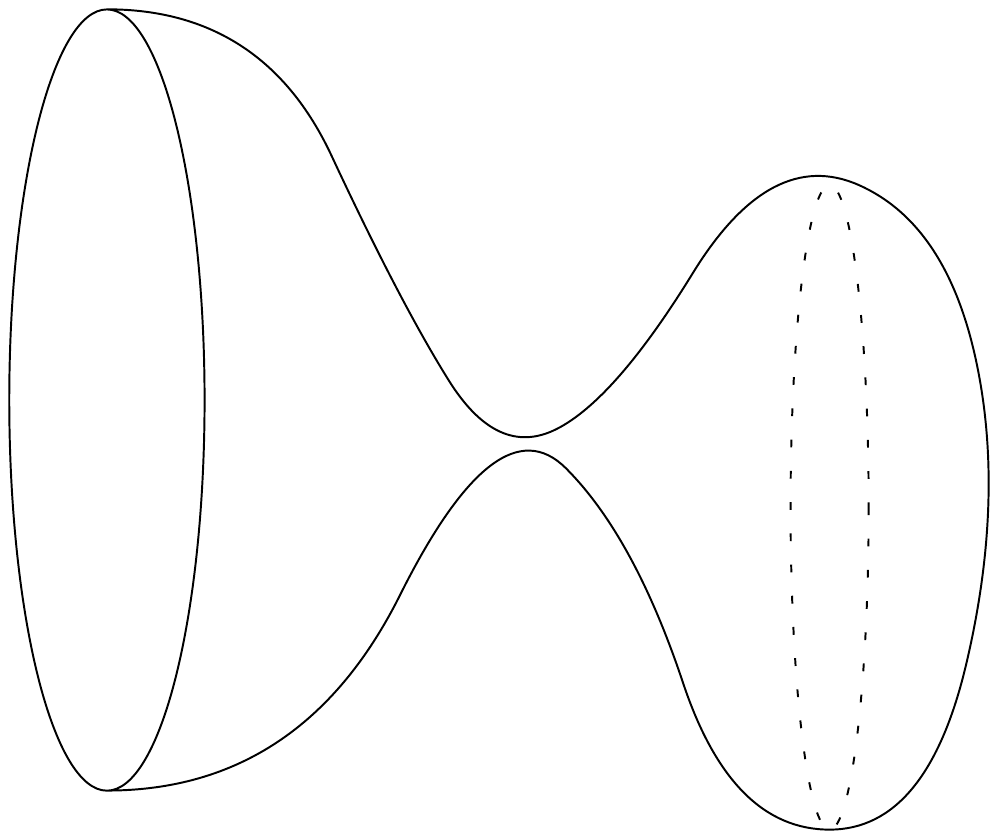}
\label{two}
\vspace{-11cm}
\caption{A typical member of a family of manifolds  developing a singularity
as
the width of the neck connecting the two parts goes to zero.}
\end{center}
\end{figure}

The degeneration of the metric (see Fig. 6),
can be obtained by considering surfaces 
(or  manifolds in the higher dimensional
cases)
with a thin ``neck'' that is pinched. At the limit the manifold contains
a pocket about which the boundary measurements
do not give any information. If the collapsing of the
manifold is done in an appropriate way, 
we have, in the limit, a singular Riemannian manifold which is
indistinguishable
in boundary measurements  from a
flat surface. 
Then the conductivity which corresponds to this
metric is also singular at the pinched points, cf. the first formula in
(\ref{eq: cond}).
The electrostatic measurements on the boundary for this
singular  conductivity
will be the same
 as for  the original regular  conductivity
corresponding to the metric $g$.

To give a  precise, and concrete, realization of this idea, 
let $B(0,R)\subset \R^3$ denote the
open ball
with center 0 and radius $R$. We use in the sequel the set $N=B(0,2)$,
the region at the boundary of which the electrostatic measurements will be
made,
decomposed into two parts, $N_1=B(0,2)\setminus \overline B(0,1)$
and  $N_2=B(0,1)$.
We call the interface $\Sigma=\p N_2$  between
$N_1$ and $N_2$ the {\it cloaking surface}.

We  also use  a ``copy'' of the ball $B(0,2)$, with the
notation $M_1=B(0,2)$,  another ball $M_2=B(0,1)$,
and the disjoint union $M$ of $M_1$ and $M_2$. (We will see the reason for
distinguishing between $N$ and $M$.)
Let $g_{jk}=\delta_{jk}$ be the Euclidian metrics  in $M_1$
and $M_2$
and let $\gamma=1$ be the corresponding isotropic 
homogeneous conductivity.
We define a singular transformation
\beq
\label{eqn-sing transf}
F_1:M_1\setminus\{0\}\to
N_1 ,\quad
   \ F_1(x)=(\frac {|x|}2+1)\frac x{|x|},\quad
0<|x|\le 2,
\eeq
and a regular transformation (diffeomorphism) $F_2: M_2 \mapsto
N_2$,
which for simplicity we take to be the identity map $F_2=Id$.
Considering the maps $F_1$ and $F_2$ together, $F= (F_1, F_2)$, 
we define a map $F:M\setminus \{0\}= (M_1\setminus
\{0\})\cup M_2
\to N\setminus \Sigma$.

The push-forward $\tilde g=F_*g$ of the metric $g$ in $M$ by $F$ is
the metric in $N$ given by
\beq\label{transf 2}
\left(F_*g\right)_{jk}(y)=\left.
\sum_{p,q=1}^n \frac {\p F^p}{\p x^j}(x)
\,\frac {\p F^q}{\p x^k}(x)  g_{pq}(x)\right|_{x=F^{-1}(y)}.
\eeq
This metric gives rise to a  conductivity $\tilde \sigma$ in $N$ which is
singular in $N_1$,
\beq\label{eq: cond}
\tilde \sigma =\left\{\begin{array}{ll}
|\tilde g|^{1/2}\tilde g^{jk}
  & \hbox{for }x\in N_1,\\
\delta^{jk} & \hbox{for }x\in N_2.\end{array}\right.
\eeq

Thus, $F$ forms an invisibility construction that
we call the ``blowing up a point". 
Denoting by
$(r,\phi,\theta)\mapsto
      (r\sin\theta \cos \phi,r\sin\theta \sin \phi,r\cos\theta)$ the
spherical coordinates,
we have
\begin{equation}\label{eqn-sing tensor 2}
\tilde \sigma=
\left(\begin{array}{ccc}
2(r-1)^2\sin \theta & 0 & 0\\
0 & 2 \sin \theta & 0 \\
0 & 0 &  2 (\sin \theta)^{-1}\\
\end{array}
\right), \quad 1<|x|\leq 2.
\end{equation}
Note that the anisotropic conductivity $\tilde \sigma$ is singular 
{degenerate}
on
$\Sigma$ in the sense that it is not bounded from below by any positive
multiple of $I$. (See \cite{KSVW} for a similar calculation.)
The Euclidian conductivity $\delta^{jk}$  in $N_2$ (\ref{eq: cond})
could be replaced by any smooth conductivity bounded from
below and above by positive constants. This would correspond
to cloaking of a general object with  non-homogeneous, anisotropic
conductivity. 
Here, we use the Euclidian
metric just for simplicity.

Consider now the {\emph{Cauchy data}} of all solutions in the Sobolev space
$H^1(N)$ of the
conductivity equation
corresponding to $\tilde \sigma$, that is,
\ba
C_1(\tilde \sigma)=\{(u|_{\p N},\nu\cdotp \tilde \sigma \nabla u|_{\p N})\
:\ u\in H^1(N),\
\nabla\cdotp \tilde \sigma\nabla u=0\},
\ea
where $\nu$ is the Euclidian unit normal vector of $\p N$.

\begin{theorem}\label{thm-cond}
(\cite{GLU3}) The Cauchy data of all
$H^1$-solutions for the conductivities $\tilde \sigma$ and $\gamma$ on $N$
coincide, that is, $C_1(\tilde \sigma)=C_1(\gamma)$.
\end{theorem}

This means that all boundary measurements for
  the homogeneous conductivity $\gamma=1$ and the degenerated
conductivity $\tilde \sigma$  are the same. The result
above was proven in \cite{GLU2,GLU3} for the case of dimension $n\ge 3.$ The
same basic construction works in the two dimensional case \cite{KSVW}. For
a further  study of  the limits of visibility and invisibility in two
dimensions,
see \cite{ALP2}.

Fig.\ 2 portrays an
analytically obtained solution on a disc with conductivity
$\tilde \sigma$.
As seen in the figure, no currents appear near the center
of the disc, so that if the conductivity is changed near
the center, the measurements on the boundary $\p N$ do not change.

The above invisibility result  is valid for a more general class of singular
cloaking transformations,
\eg, quadratic singular
transformations for Maxwell's equations which were introduced first in \cite
{Cai2}.
A general class, sufficing at least for
electrostatics, is given by the following result from
\cite{GLU3}:

\begin{theorem}\label{main B}
Let $\Omega\subset \R^n$, $n\geq 3$, and $g=(g_{ij})$ a smooth metric on $\Omega$
bounded from above and below by positive constants.
Let $D\subset\subset \Omega$ be such there is
a $C^\infty$-diffeomorphism $F:\Omega\setminus\{y\}\to
\Omega\setminus \overline D$ satisfying
$F|_{\p \Omega}=Id$ and such that
\beq\label{Q 4}
dF(x)\geq c_0I,\quad
\hbox{det}\,(dF(x))\geq c_1\,\hbox{dist}_{{}_{\R^n}}\,(x,y)^{-1}
\eeq
where $dF$ is the Jacobian matrix in Euclidian coordinates
on $\R^n$ and $c_0,c_1>0$.
Let $\hat g$ be a metric in $\Omega$ which coincides with
$\tilde g=F_*g$ in $\Omega\setminus \overline D$
and is an arbitrary regular positive definite metric
in $D^{int}$.
 Finally, let  $\sigma$ and $\hat \sigma$
be the conductivities corresponding to $g$ and $\hat g$,
cf.\ (\ref{eqn1.13}).
Then,
\ba
C_1(\hat \sigma)=C_1(\sigma).
\ea
\end{theorem}

The key to the proof of Thm. \ref{main B} is a removable singularities
theorem that
implies that solutions of the conductivity equation in  $\Omega\setminus
\overline D$ 
pull back
by this singular transformation to solutions of
the conductivity equation in the whole $\Omega$.

Returning to the case $\Omega=N$ and the conductivity given by 
(\ref{eq: cond}),
similar type of results are valid also for a more
general class of solutions.
Consider an unbounded quadratic form,
$A$ in $L^2(N, |\tilde g|^{1/2} dx)$,
\ba
A_{\tilde \sigma}[u,v]=\int_N\tilde\sigma\nabla u\cdotp \nabla v\,dx
\ea
defined for $u,v\in {\mathcal D}(A_{\tilde \sigma})=C_0^\infty(N)$.
Let $\overline A_{\tilde \sigma}$ be the closure of this
quadratic form and say that

\ba
\nabla\cdotp \tilde \sigma\nabla u=0\quad\hbox{in }N
\ea
is satisfied in the finite energy sense if there is
$u_0\in H^1(N)$ supported in $N_1$
such that $u-u_0\in {\mathcal D}(\overline A_{\tilde \sigma})$ and
\ba
\overline A_{\tilde \sigma}[u-u_0,v]=
-\int_N\tilde\sigma\nabla u_0\cdotp \nabla v\,dx,\quad \hbox{for all }v\in
{\mathcal D}(\overline A_{\tilde \sigma}).
\ea
Then the Cauchy data set of the finite energy solutions, denoted by
\ba
C_{f.e.}(\tilde \sigma)=\Big\{(u|_{\p N},\nu\cdotp \tilde \sigma \nabla u|_{\p
N})\
:\
\hbox{$u$ is a finite energy solution of
$\nabla\cdotp \tilde \sigma\nabla u=0$}\Big\}
\ea
coincides with the Cauchy data $C_{f.e.}(\gamma)$ corresponding
to the homogeneous conductivity $\gamma=1$, that is,
\beq\label{fin. en. solutions}
C_{f.e.}(\tilde \sigma)=C_{f.e.}(\gamma).
\eeq
This and analogous results for the corresponding
equation in the
non-zero
frequency case,
$$
\nabla \cdot \tilde \sigma \nabla u = \lambda u,
$$
were considered in \cite{GKLU1}.   We will discuss them in more detail 
 in the next section.

We emphasize that the above results were obtained in dimensions $n\ge 3$.
Kohn, Shen, Vogelius and Weinstein \cite{KSVW} have shown that the
singular conductivity resulting from the same transformation also cloaks for
electrostatics in two dimensions.

\section{Optical Invisibility:  \\ Cloaking at Positive
Frequencies}\label{sec-2006}

\subsection{Developments in physics}

Two  transformation optics--based 
invisibility cloaking constructions were proposed 
in 2006 \cite{Le, PSS1}. Both of these were expressed in the \emph{frequency
domain}, \ie, for monochromatic waves. Even though 
the mathematical models can be considered at any frequency,
it is important to note that the custom designed {\it metamaterials}
manufactured for physical implementation
of these or similar designs
are very dispersive; that is, the relevant material parameters (index of
refraction, etc.) depend on the frequency. Thus, physical cloaking
constructions
with current technology are  essentially monochromatic, working over at best
a very narrow range of frequencies. The many interesting issues in physics
and engineering that this difficulty raises are beyond the scope of this
article; see \cite{NJP} for recent work in this area.

Thus, we will also work in the frequency domain and will be interested in
either scalar waves of the form $U(x,t)=
u(x)e^{ikt}$, with $u$ satisfying the Helmholtz equation,
\bq\label{eqn-Helmholtz}
(\D+k^2n^2(x))u(x)=\rho(x),
\eq
where $\rho(x)$ represents any internal source present, or in
time-harmonic
electric and magnetic fields $\bE(x,t)=E(x)e^{ikt},\,
\bH(x,t)=H(x)e^{ikt}$, with $E,H$ satisfying Maxwell's equations,
\bq\label{eqn-Maxwell}
\nabla\times H = -ik \e E +J,\quad \nabla\times  E=ik\mu H,
\eq
where $J$ denotes any external current present.

To review the ideas of \cite{PSS1} for  electromagnetic cloaking
construction,
let us start with Maxwell's equations in three dimensions.
We consider a ball $B(0,2)$ with
the homogeneous, isotropic material parameters, the
permittivity $\e_0\equiv1$ and the permeability $\mu_0\equiv1$. 
Note that, with respect to a smooth coordinate transformation,
the permittivity and the permeability transform in the same way
(\ref{eqn1.6}) as conductivity.
Thus, pushing  $\e_0$ and $\mu_0$ forward
by the ``blowing up a point"
map $F_1$ introduced in (\ref{eqn-sing transf})
yields permittivity $\tilde \e(x)$
and permeability $\tilde \mu(x)$ which are  inhomogeneous and
anisotropic. 
In  spherical coordinates,  the
representations of $\tilde \e(x)$ and $\tilde \mu(x)$ 
are identical to the conductivity $\tilde \sigma$ 
given in 
 (\ref{eqn-sing tensor 2}).
They are are smooth and non-singular in the open domain
$N_1:=B(0,2)\setminus \overline B(0,1)$ but, as seen from
(\ref{eqn-sing tensor 2}), degenerate  as $|x|\lra 1^+$, i.e. at the
  cloaking surface $\Sigma=\{|x|=1\}$. One of the eigenvalues, namely the
  one 
associated with the radial direction, behaves as $2(|x|^2-1)^2$ and tends to
zero as $|x| \to 1^+$.
This determines the electromagnetic parameters in the image of $F_1$,
that is, in $N_1$. In $N_2$ we can choose
the electromagnetic parameters $\e(x),\mu(x)$ to be any  smooth, nonsingular
tensors. The material parameters in $N_2$ correspond to 
an arbitrary object
we want to hide from exterior measurements.

In the following, we refer to  $N:=N_1\cup N_2\cup \Sigma=B(0,2)$
with the described material parameters
as the \emph{cloaking device}
and denote the resulting specification of the material parameters
on $N$  by
$\te,\tm$.
As noted, the representations of $\te$ and $\tm$
on $N_1$
coincide with that of $\tilde \sigma$ given by (\ref{eq: cond})  in
spherical coordinates.
Later, we will also describe the
\emph{double coating} construction, which corresponds to appropriately
matched layers of metamaterials on both the outside and the inside of
$\Sigma$.

The construction above is  what we call the \emph{single coating} 
 \cite{GKLU1}. This theoretical description of an invisibility device can,
 in principle, be physically realized 
by taking an arbitrary object in
$N_2$ and surrounding it with special material, 
located in $N_1$, which implements the values of $\tilde{\e},\tilde\mu$. 
Materials with customized values of $\epsilon$ and $\mu$ (or other material
parameters) are referred to as {\it metamaterials}, the study of which has
undergone an explosive growth in recent years. {There is no universally accepted definition of metamaterials, which seem to be in the ``know it when you see it" category. However, the label usually attaches 
to  macroscopic material structures having a manmade  one-, two- or
three-dimensional cellular architecture,  and producing combinations of material parameters not available in nature (or
even in conventional composite materials), due to resonances induced by the geometry of the cells  \cite{walser,Elef}. }
Using metamaterial cells (or ``atoms", as they are sometimes called),
designed to resonate at the desired frequency, it is possible to specify the
permittivity and permeability tensors
fairly arbitrarily \emph{at a given frequency}, so that they may have very large,   very small or even negative eigenvalues.  The use of resonance phenomenon
also  explains why
the material properties of metamaterials  strongly depend
on the  frequency, and broadband metamaterials may not be possible.}

\subsection{Physical justification of cloaking}

To understand the physical arguments  describing the behavior of
electromagnetic waves in the cloaking device,
consider Maxwell's equations exclusively 
on the open annulus $N_1$ and in the punctured ball $M_1\setminus \{0\}$.
Between these domains, the  transformation
$F_1:M_1\setminus \{0\}\to N_1$ is smooth. Assume that
the electric field $E$ and the magnetic field $H$ 
in $M_1\setminus \{0\}$ solve  Maxwell's equations,
\bq\label{eqn-Maxwell I}
\nabla\times H = -ik \e_0 E,\quad \nabla\times  E=ik\mu_0 H
\eq
with constant, isotropic $\e_0,\, \mu_0$.
Considering $E$ as a differential 
1-form $E(x)=E_1(x)dx^1+E_2(x)dx^2+E_3(x)dx^3$
we define the push-forward of $E$ by $F_1$, denoted $\tilde E=(F_1)_* E$,
in $N_1$, by
\ba
\tilde E(\tilde x)=\sum_{j=1}^3\tilde E_j(\tilde x)d\tilde x^j
=
\sum_{j=1}^3
\bigg (\sum_{k=1}^3(DF^{-1})_j^k(\tilde
x)\,E_k(F^{-1}(\tilde
x))\bigg) d\tilde x^j,\quad \tilde x=F(x).
\ea
Similarly, for the magnetic field $H$ we define
$\tilde H=(F_1)_*H$ in $N_1$. Then $\tilde E$ and $\tilde H$ satisfy
Maxwell's equations in $N_1$,
\bq\label{eqn-Maxwell II} 
\nabla\times \tilde H = -ik  \tilde \e  \tilde E,
\quad \nabla\times  \tilde E=ik \tilde\mu  \tilde H,
\eq
where the material parameters in $\tilde \e, \tilde\mu$ are defined {in $N_1$} by 
\ba
\tilde \e=(F_1)_*\e_0=\tilde \sigma,\quad
\tilde \mu=(F_1)_*\mu_0=\tilde \sigma.
\ea
Here $\tilde \sigma$ is given by (\ref{eqn-sing tensor 2}).
  
Thus, the solutions $(E,H)$ in
the open annulus $N_1$ and solutions $(\tilde E,\tilde H)$ 
in the punctured ball $M_1\setminus \{0\}$
are in a one-to-one correspondence. If one compares
just the solutions  in these domains, {without considering the
behavior within the cloaked region $N_2$ or }any boundary
condition on the cloaking surface $\Sigma$,
the observations of the possible solutions of Maxwell's equations at $\p
N=\p B(0,2)$
are unable to distinguish between the
cloaking device $N$, with an object hidden from view in $N_2$,
and the empty
space  $M$.

{One can also consider the behavior of light rays, corresponding 
to the high frequency limits of solutions; see also \cite{Le}, which
proposed cloaking for scalar optics in $\R^2$.} These are, mathematically
speaking, the geodesics 
on the manifolds $(M_1,g)$ and $(N_1,\tilde g)$, see Fig.\ 7. One observes that
almost all geodesics $\mu$ on $N_1$ don't hit the cloaking surface $\Sigma$
but go around the domain $(N_2,\tilde g)$ and have the same intrinsic
lengths (\ie,
travel times) as the corresponding geodesics $\tilde \mu=F_1^{-1}(\mu)$ on
$(M_1,g)$.
Thus, roughly speaking, almost all light rays sent into $N_1$ from $\p N$
go around  the ``hole"  $N_2$,
and reach $\p N$ in the same time as the corresponding  rays on $M$.

\begin{figure}[htbp]\label{schematic rays}
\begin{center}
\includegraphics[width=.45\linewidth]{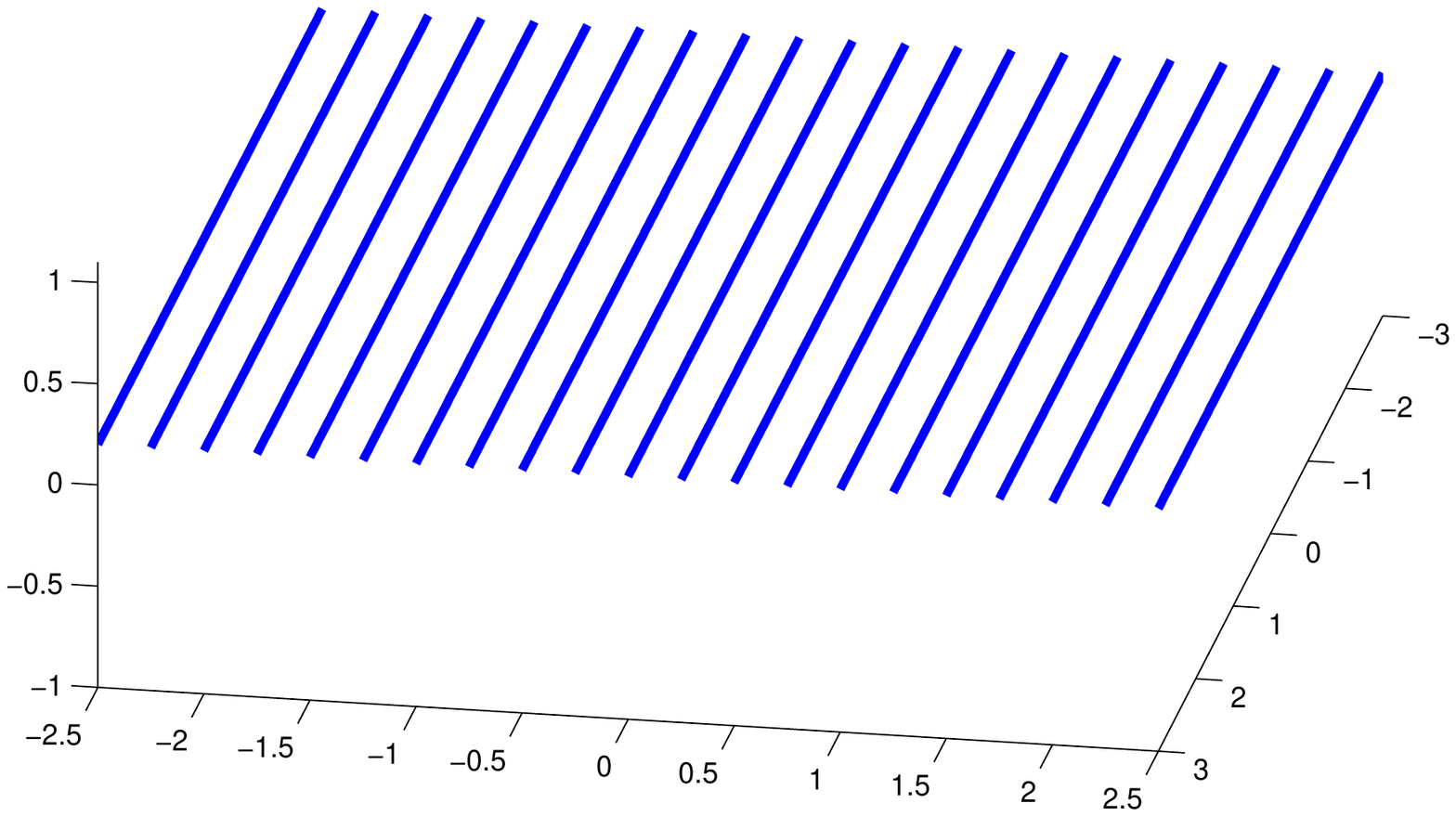}\quad \includegraphics[width=.45\linewidth]{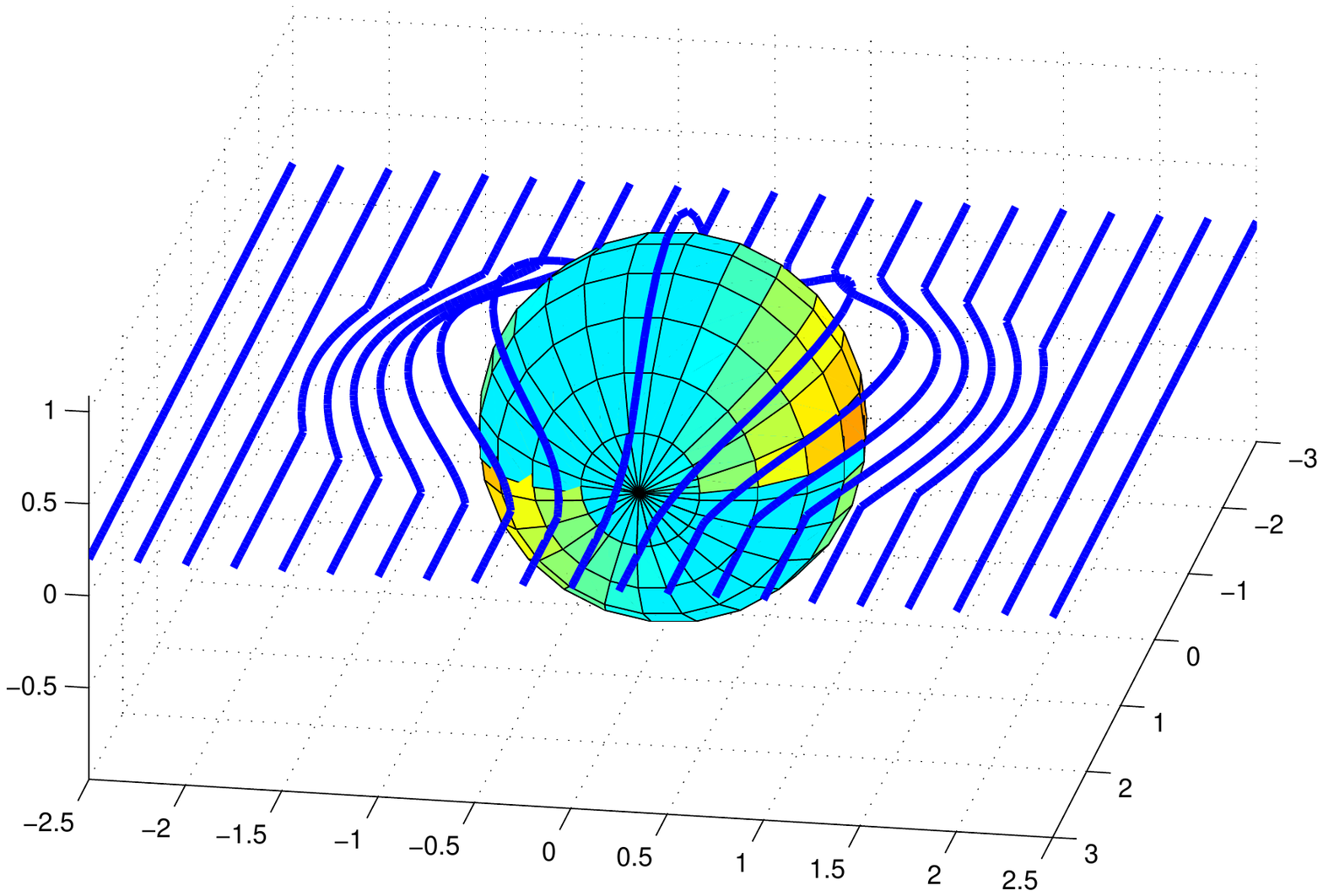}
\caption{Left, light rays are shown in the Euclidian space $\R^3$ and
right, the same light rays are shown when a cloaking device  $(N,\tilde g)$ is located
in the ball $B(0,1)$. The metamaterial in which the light rays travel is not shown; the
sphere is the cloaking surface $\p B(0,1)$. On left, the light rays correspond to geodesics   on 
$(M_1\setminus \{0\},g)$  and on right, the geodesics on $(N_1,\tilde g)$. The map
$F_1$ maps the geodesics on $M_1$ (not passing through origin) to those of
$N_1$.
}
\end{center}
\end{figure}

{The cloaking effect was justified  in
\cite{PSS1} on the level of the chain rule for $F_1$, and in the sequels
\cite{PSS2,CPSSP} on the level of rays and numerical simulations, on $N_1$. 
We will see below that studying  the  behavior
of the  waves on the \emph{entire} space, including in the cloaked region
$N_2$ and at the cloaking surface $\Sigma$, is crucial to fully
understanding cloaking and its limitations.}

A particular difficulty is that, due to the
degeneracy of
$\te$ and
$\tm$, the weighted
$L^2$ space defined by the energy norm
\bq\label{eqn-energy}
\|\tilde E\|_{L^2(N, |\tg|^\frac12 dx)}^2+\|\tilde H\|_{L^2(N,
|\tg|^\frac12 dx)}^2=
\int_{N}
(\tilde \e^{jk}\, \tilde E_j\, \overline{  \tilde E_k}
+
\tilde \mu^{jk}\, \tilde H_j\, \overline{ \tilde H_k}) \,dx
\eq
includes 
forms, which are not distributions, \ie, not
in the dual of the vector fields having $C^\infty_0(N)$ coefficients.
Indeed, this class contains the forms with the radial component 
behaving like
$O((r-1)^{-\a})$ in the domain $r>1$, where $1<\a <3/2$.
The meaning of  the Helmholtz or Maxwell's equations for such ``waves"  is
problematic, and to treat cloaking
rigorously, one should consider the boundary measurements (or scattering
data) of finite energy waves which also satisfy Maxwell's equations in some
reasonable weak sense.
Analysis of cloaking from this more rigorous point of view was carried out
in
\cite{GKLU1}, which forms the basis for much
of the discussion here.

\subsection{Cloaking for the Helmholtz equation}\label{subsec-Helmholtz}

Let us start with the cases of scalar optics or acoustics, governed in the
case of isotropic media by the Helmholtz equation (\ref{eqn-Helmholtz}). In
order to work with anisotropic media, we convert this to the Helmholtz
equation with respect to a Riemannian metric $g$. Working in dimensions
$n\ge 3$, we take advantage of the  one-to-one correspondence
(\ref{eqn1.13}) between (positive definite) {
conductivities} and Riemannian metrics $g$.
{
Let us consider the Helmholtz equation
\bq\label{eqn-metric Helmholtz}
(\D_g+k^2)u=\rho,
\eq
where $\D_g$ is the Laplace-Beltrami operator associated with the Euclidian
metric $g_{ij}=\delta_{ij}$.} Under a smooth diffeomorphism
$F$, the metric
$g$ pushes forward to a metric
$\tg=F_*g$, and then, for $u= \tu\circ F$, we
have
\beq\label{eq: Change of coordinates}
(\D_g+k^2)u=\rho \iff (\D_{\tg}+k^2)\tu=\tr,
\eeq                 
where  $\rho=\tr\circ F$.

Next we consider the case when $F(x)$ is not a smooth diffeomorphism,
but the one introduced by (\ref{eqn-sing transf}), if 
$x\in M_1\setminus \{0\}$ and identity, if  $x\in M_2$.

Let $\tilde f\in L^2(N,dx)$ be a function such that $\supp (\tilde
f)\cap \Sigma=\emptyset$. We now give  the precise definition of a finite
energy   solution for the Helmholtz equation.

\begin{definition}\label{energysolution}
{Let $g$ be the Euclidian metric on $M$
and $\tilde g=F_*g$ be the singular
metric on $N\setminus \Sigma$.}
A measurable function $\tilde u$ on $N$ is a
\emph{finite energy}  solution of the Dirichlet problem for
the Helmholtz
equation on
$N$,
\beq\label{case 3 B 1}
& &(\Delta_{\tilde g}+k^2)\tilde u=\tilde f\quad\hbox{on }N,\\
& &\tilde u|_{\p N}=\tilde h,\nonumber
\eeq
if
\beq\label{cond 1 B}
& &\tilde u\in L^2(N,\, |\tilde g|^{1/2}dx);\\
\label{cond 2 B}
& &\tilde u|_{N\setminus \Sigma}\in H^1_{loc}(N\setminus \Sigma,dx);
\\
\label{cond 3 B}
& &\int_{N\setminus\Sigma} |\tilde g|^{1/2}\tilde g^{ij}\p_i\tilde
u\p_j\tilde u\,dx<\infty,\\
& &\tilde u|_{\p N}=\tilde h\nonumber;
\eeq
and, for all ${\tilde \psi} \in C^\infty(N)\hbox{ with } {\tilde
\psi}|_{\p N}=0$,
\beq\label{cond 5 B}
& &\int_{N} [-(D^j_{\tilde g}\tilde u) \p_j {\tilde \psi}+k^2 \tilde
u {\tilde \psi}|\tilde g|^{1/2}]dx
=\int_{N} \tilde f(x) {\tilde \psi} (x)
|\tilde g|^{1/2} dx
\eeq
where $D^j_{\tilde g}\tilde u=
|\tilde g|^{1/2}\tilde g^{ij}\p_i u$ is defined as a Borel measure
defining a distribution on $N$.
\end{definition}

Note that the inhomogeneity $\tilde f$ is allowed to have two
components, $\tilde f_1$ and $\tilde f_2$, supported in the interiors of
$N_1,N_2$, resp.
The latter corresponds to an active object being rendered undetectable
within the cloaked region. On the other hand, the former
corresponds to an active object embedded within the metamaterial cloak
itself, whose position apparently shifts in a predictable manner according
to the transformation $F_1$; this phenomenon, which also holds for both
spherical and cylindrical cloaking for Maxwell's equations,  was later
described and numerically modelled in the cylindrical setting,  and termed
the ``mirage effect" \cite{Z}.
\bigskip

Next we consider the relation 
between the finite energy solutions on $N$ and the solutions on 
$M$.

\begin{theorem}(\cite{GKLU1})\label{single coating with Helmholtz}
Let $u=(u_1, u_2) : (M_1\setminus \{0\}) \cup M_2\to \R$
and  $\tilde u: N \setminus \Sigma\to \R$ be  measurable functions
such that $u=\tilde u\circ F$.
Let $f=(f_1, f_2):  (M_1\setminus \{0\}) \cup M_2\to \R$
and  $\tilde f:N\setminus \Sigma\to \R$ be $L^2$ functions,
supported away from $0 \in M_1$ and $\Sigma \subset N$
such that  $f=\tilde f\circ F$. At last, let
 $\tilde h:\p N\to\R,\, h:\p M_1\to\R$ be such that $h=\tilde
h\circ F_1$.

Then the following are equivalent:
\begin{enumerate}

\item The function
$\tilde u$, considered as a measurable function on $N$,
is a finite energy solution to the Helmholtz equation (\ref{case 3 B 1})
with inhomogeneity $\tilde f$ and Dirichlet data $\tilde  h$ in the sense of
Definition
\ref{energysolution}.

\item The function $u$ satisfies
\beq\label{eq on M1}
(\Delta_{g}+k^2)u_1=f_1\quad\hbox{ on }M_1,\quad
u_1|_{\p M_1 }=h,
\eeq
and
\beq\label{eq on M2}
& &(\Delta_{g}+k^2)u_2=f_2\quad\hbox{ on } M_2, \quad
g^{jk}\nu_j\p_k u_2|_{\p M_2}=b,
\eeq
with $b=0$. Here $u_1$ denotes the continuous extension of $u_1$ from
$M_1 \setminus \{0\}$ to $M_1$.
\end{enumerate}

Moreover, if $u$ solves (\ref{eq on M1}) and  (\ref{eq on M2})  with
$b\not =0$, then the function
$\tilde u=u\circ F^{-1}:N\setminus \Sigma\to \R$,
considered as a measurable function on $N$,
is not a finite energy solution to the  Helmholtz equation.
\end{theorem}

As mentioned in \S\ref{sec-intro}, and detailed in \cite{GKLU5}, this
result also describes a structure which cloaks both passive
objects and active
sources for acoustic waves.   Equivalent structures in the spherically
symmetric
case and with only  cloaking of  passive objects verified were considered
later
in \cite{Ch3,Cu}.

{The idea of the proof of Thm. \ref{single coating with Helmholtz} 
is to consider $F_1$ and $F_2$  as coordinate transformations. 
As in formula (\ref{eq: Change of coordinates}),
we see that if $u$ is a finite energy solution of the Helmholtz equation   (\ref{case 3 B 1}) in $N$
 then $u_1=u\circ F_1$, defined in $M_1\setminus \{0\}$, satisfies
 the Helmholtz equation (\ref{eq on M1}) on the set $M_1\setminus \{0\}$. Moreover, as 
the energy is invariant under a change of coordinates, one sees that $u|_{M_1\setminus \{0\}}$
is in the Sobolev space $H^1(M_1\setminus \{0\})$. Since the point $\{0\}$ 
has  Hausdorff dimension less or equal the dimension of $\R^3$ minus two,
the possible singularity of $u_1$ at zero is removable (see \eg, \cite{KKM}), that is, $u_1$ 
has an extension to a function defined on the whole ball $M_1$ so that the Helmholtz equation 
(\ref{eq on M1}) is satisfied on the whole ball.

Let us next discuss the appearance of the Neumann boundary condition in (\ref{eq on M2}).
Observe that in Def. \ref{energysolution} the Borel measure $D^j_{\tilde g}v=
|\tilde g|^{1/2}\tilde g^{ij}\p_i v$ is absolutely continuous with respect to
the Lebesgue measure for all functions $v\in C^\infty_0(N)$. We can approximate
the finite energy solution $\tilde u$ of equation (\ref{case 3 B 1}) with source $\tilde f$, 
supported away from $\Sigma$, by such functions. This yields that the measure of the cloaking surface satisfies
$D^j_{\tilde g}\tilde u(\Sigma)=0$. Thus, using integration by parts, we see for arbitrary 
$\tilde \psi \in C^\infty_0(N)$ that
\beq\nonumber
0&=&\lim_{\e\to 0+}\int_{B(0,1+\e)\setminus B(0,1-\e)} 
[(D^j_{\tilde g}\tilde u) \p_j {\tilde \psi}-k^2 \tilde
u {\tilde \psi}|\tilde g|^{1/2}]dx\\ \label{cond 5 B vers2}
&=&\lim_{\e\to 0+}\left(\int_{\p B(0,1+\e)}-\int_{\p B(0,1-\e)}\right) 
[\nu_j\,(|\tilde g|^{1/2}\tilde g^{ij}\p_i \tilde u] \,  {\tilde \psi}\,dS(x),
\eeq
where $dS$ is the Euclidian surface area. Changing coordinates by 
$F_1^{-1}:\p B(0,1+\e)\to \p B(0,2\e)$
in the first integral in (\ref{cond 5 B vers2}) and letting
$\e\to 0$  in the second integral, we see that
\beq\label{eq: boundary integral}
0=\lim_{\e\to 0+}\int_{\p B(0,2\e)}\frac {\p u_1}{\p \nu} \psi \, dS- 
\int_{\Sigma} \left. \frac {\p \tilde u}{\p \nu}\right|_{\Sigma-}\tilde \psi\,dS,
\eeq
where $\psi=(F_1)^*\tilde \psi$ is a bounded function on $M_1$
and $u_1$ is the solution of (\ref{eq on M1}) in $M_1$, hence 
smooth  near $0$. Here $\left.\frac {\p u}{\p \nu}\right|_{\Sigma-}$ denotes
the interior normal derivative. Thus, the first integral in (\ref{eq: boundary integral})
over the sphere of radius $2\e$ goes to zero as $\e\to 0$ yielding 
that the last integral must vanish. As $\tilde \psi$ is arbitrary,
this implies that $u$  satisfies the homogeneous boundary condition
on the inside of the cloaking
surface $\Sigma$.  
We point out that this Neumann boundary condition} is a consequence
of the fact that the coordinate transformation $F$ is  singular
on the cloaking surface $\Sigma$. See also \cite{talkVogelius} for the planar case.

\subsection{Cloaking for Maxwell's equations}\label{subsec-Maxwell}

In what follows, we treat Maxwell's equations in non-conducting and lossless
media,
that is, for which the conductivity vanishes 
and the components of $\e,\mu$ are real
valued.
Although somewhat suspect (presently, metamaterials are quite lossy), these
are standard assumptions in the physical literature. We point out that Ola,
{P\"aiv\"arinta}
and Somersalo \cite{OPS} have shown that cloaking is not possible
for Maxwell's equations with non-degenerate \emph{isotropic}, sufficiently smooth,
electromagnetic
parameters.

We will use the invariant formulation of Maxwell's
equations. To this end, consider
a smooth compact oriented connected Riemannian 3-manifold $M$, $\partial
M\neq
\emptyset$,
with a metric $g$, that we call the background metric.
Clearly, in physical applications we take $M\subset \R^3$ with $g$
being the Euclidean metric $g_0$.
Time-harmonic Maxwell's equations on the manifold $M$
are  equations of the form
\beq
\label{Maxwell-Faraday}
& & {\rm curl}\,E(x) = ik B(x),\\
\label{Maxwell--Ampere}
& & {\rm curl }\,H(x) =-ik D(x)+J.
\eeq
Here the electric field  $E$ and the magnetic field 
$H$ are 1-forms and the electric flux $D$ and the magnetic flux $B$ are
 2-forms, and curl is the standard exterior differential $d$.
The external current $J$ is considered also as a 2-form.
The above fields are related by the constitutive relations,
\beq
\label{constitutive}
D(x) = \varepsilon(x)E(x),\quad
 B(x) = \mu(x)H(x),
\eeq
where $\varepsilon$ and $\mu$ are linear maps from
1-forms to 2-forms. 
Thus, in local coordinates on $M$, {we denote
\ba
E=E_j(x)dx^ j,\quad 
D=D^1(x)dx^2\wedge dx^3+
D^2(x)dx^3\wedge dx^1+D^3(x)dx^1\wedge dx^2,\\
H=H_j(x)dx^ j,\quad
B=B^1(x)dx^2\wedge dx^3+
B^2(x)dx^3\wedge dx^1+B^3(x)dx^1\wedge dx^2.
\ea
Using these notations, the constitutive relations
take the form $D^j=\e^{jk}E_k$ and $B^j=\mu^{jk}H_k$.}

Note, that in the case of a homogeneous  Euclidian space,
where $\e_0=1, \mu_0=1$, {the operators
$\e$ and $\mu$ correspond to
the standard Hodge star operator
$*:\Omega^1(\R^3)\to \Omega^2(\R^3)$ corresponding to the
Euclidian metric $(g_0)_{jk}=\delta_{jk}$.}
On an arbitrary manifold $(M, g)$ it is always possible
to define the permittivity $\e$  and  permeability $\mu$,
{to be the Hodge star operator corresponding
to the metric $g$.}
Then, in local coordinates on $M$, 
\beq \label{63a}
\e^{jk}=\mu^{jk}=
| g|^{1/2} g^{jk}.
\eeq
This type of electromagnetic material parameters,
which has the same transformation law, under the change of coordinates,
as the conductivity, was studied in \cite{KLS}.

To introduce the material parameters $\tilde \e(x)$ and  $\tilde \mu(x)$
in the ball $N=B(0,2)\subset \R^3$
that make cloaking possible,  we start with the singular map $F_1$ given by
(\ref{eqn-sing transf}). We then introduce the Euclidean metric  on
$N_2$
and the metric $\tilde g=F_*g$ in $N_1$. 
{Finally,
we define the singular permittivity and permeability 
in $N$ using the transformation rules (\ref{63a}) which lead to}
the
formulae analogous to (\ref{eq: cond}),
\beq\label{def: tilde e and mu}
\tilde \e^{jk}=\tilde \mu^{jk}=
\left\{\begin{array}{ll}
|\tilde g|^{1/2}\tilde g^{jk}
  & \hbox{for }x\in N_1,\\
\delta^{jk} & \hbox{for }x\in N_2.\end{array}\right.
\eeq
Clearly that, as in the case of Helmholtz equations, these material
parameters are singular
on $\Sigma$. 

We note that in $N_2$ one could define $\tilde \e$ and $\tilde \mu$ to be
arbitrary smooth non-degenerate material parameters. For simplicity, we
consider
here only the homogeneous material in the cloaked region $N_2$.

\subsection{Definition of
solutions of Maxwell equations}

{In the rest of this section, $\e=1$ and $\mu=1$ on the manifold $M$ and
$\tilde \e $ and $\tilde \mu$ are singular material parameter on $N$
defined in (\ref{def: tilde e and mu}).
}

Since the material parameters $\tilde \e$ and $\tilde \mu$
are again singular at the cloaking surface $\Sigma$,
we need a careful formulation of the notion of a
solution.

\begin{definition}\label{Maxwell-def}
We say that  $(\tilde E, \tilde H)$ is a finite energy solution
to Maxwell's equations on $N$,
\begin{equation}\label{eqn-4.1-main}
\nabla\times \tilde E = ik \tilde \mu(x)  \tilde H,\quad \nabla\times
              \tilde H =-ik  \tilde \e(x) \tilde E+\tilde J\quad
\hbox{ on }N,
\end{equation}
if $ \tilde E$, $\tilde H$ are
one-forms and $\tilde D:=\tilde \e\, \tilde E$ and
    $\tilde B:=\tilde \mu\, \tilde H$ two-forms in $N$ with
$L^1(N,dx)$-coefficients satisfying
\beq\label{eq: Max norm1}
\|\tilde E\|_{L^2(N,|\tilde g|^{1/2}dV_0(x))}^2= \int_{N}
\tilde \e^{jk}\, \tilde E_j\, \overline{  \tilde E_k} \,dV_0(x)<\infty
,\\ \label{eq: Max norm2}
\|\tilde H\|_{L^2(N,|\tilde g|^{1/2}dV_0(x))}^2=\int_{N}
\tilde \mu^{jk}\, \tilde H_j\, \overline{ \tilde H_k} \,dV_0(x)<\infty;
\eeq
where $dV_0$ is the standard Euclidean volume and
\beq\label{eq: weak condition}
& &\int_N ((\nabla\times \tilde h)\,\cdotp \tilde E-
              ik \tilde h \,\cdotp \tilde \mu(x)\tilde H) \,dV_0(x)=0,\\
& & \nonumber
\int_N (
(\nabla\times \tilde e)\,\cdotp \tilde H+
              \tilde e \,\cdotp ( ik\tilde \e(x)\tilde E-\tilde J))
\,dV_0(x)=0
\eeq
for all 1-forms $\tilde e,\tilde h$ on $N$
having in the Euclidian coordinates components  in $C^\infty_0(N)$.
\end{definition}

Above, the inner product ``$\cdotp$'' denotes the Euclidean inner
product. {We emphasize that in Def. \ref{Maxwell-def} we assume
that the components of the physical fields $\tilde E,\tilde H,\tilde B,$ and
$\tilde D$
are integrable functions. This in particular implies that the components
of these fields are distributions.
Note that the map $F_*$ does
not map distributions on $M$
 isomorphically to distributions on $N$.
This is because $F^*:\phi\mapsto \phi\circ F$ 
does not map $C^\infty_0(N)$ to $C_0^\infty(M)$. 
Hence, on $M$ there are currents (i.e. sources) $J$, whose support
contains the point zero that do not correspond to distributional
sources $\tilde J$ on $N$ {for which $\tilde J=F_*J$ in $N\setminus
\Sigma$}. Below we will show that in the case
when a source $J$ is not supported on $N_2\cup \Sigma$,
there exist solutions for Maxwell's equations on $N$ with the 
corresponding source so that
$\tilde J=F_*J$ in $N\setminus \Sigma$}. Also, we show that
surprisingly, the finite energy solutions do not exist for
generic currents $\tilde J$. Roughly speaking, the fact that the map $F$ can
not be extended to the whole $M$, so that it would map the 
differentiable structure on $M$ to that of $N$, seems to
be the reason for this phenomena.

{Below, we denote $M\setminus \{0\}=
(M_1\setminus \{0\})\cup M_2$.}

\begin{theorem}(\cite{GKLU1})\label{single coating with Maxwell A}
Let $E$ and $H$ be 1-forms
with measurable coefficients
on $M\setminus \{ 0\}$ and $\tilde E$ and $\tilde H$ be
             1-forms
with measurable coefficients on $N\setminus \Sigma$
such that $\tilde E=F_*E$, $\tilde H=F_* H$.
Let $J$ and $\tilde J$ be  2-forms
with smooth coefficients on $M\setminus \{ 0\}$ and
            $N\setminus \Sigma$, that are supported away
from $\{ 0\}$ and $\Sigma$ such that $\tilde J=F_*J$.

Then the following are equivalent:
\begin{enumerate}
\item
The 1-forms $ \tilde E$ and $ \tilde H$ on $N$ satisfy Maxwell's equations
\beq\label{eq: physical Max M}
& &\nabla\times \tilde E = ik \tilde \mu(x)  \tilde H,\quad \nabla\times
             \tilde H =-ik  \tilde \e(x) \tilde E+\tilde J\quad
\hbox{ on }N,
\\
& &\nonumber \nu\times \tilde E|_{\p N}=f
\eeq
in the sense of Definition \ref{Maxwell-def}.

\item
The forms $E$ and $H$ satisfy Maxwell's equations on $M$,
\beq\label{eq: physical Max single M1 new}
& &\nabla\times   E = ik   \mu(x)    H,\quad \nabla\times
              H =-ik    \e(x)   E+  J\quad
\hbox{ on }M_1,
\\
& &\nonumber \nu\times E|_{\p M_1}=f
\eeq
and
\beq\label{eq: physical Max single M2 new}
& &\nabla\times   E = ik   \mu(x)    H,\quad \nabla\times
              H =-ik    \e(x)   E+  J\quad
\hbox{ on }M_2
\eeq
with Cauchy data
\beq\label{eq: physical Max single M3 new}
& &\nu\times   E|_{\p M_2}=b^e,\quad
\nu\times   H|_{\p M_2}=b^h
\eeq
that satisfies $b^e=b^h=0$.
\end{enumerate}

Moreover, if $E$ and $H$ solve (\ref{eq: physical Max single M1 new}),
(\ref{eq: physical Max single M2 new}), and (\ref{eq: physical Max single M3
new})
with non-zero $b^e$ or $b^h$, then the fields
$\tilde E$ and $\tilde H$ are not solutions of Maxwell equations
on $N$ in the sense of Definition \ref{Maxwell-def}.
\end{theorem}

{Let us briefly discuss the proof of this theorem. 
In  Euclidian space, with $\e=1$ and $\mu=1$, Maxwell's
equations (\ref{eqn-Maxwell}) with $J=0$ and $k\not =0$ imply that  the divergence
of $D$ and $B$ fields are zero, or equivalently that
\ba
\nabla\,\cdotp (\e E)=0,\quad \nabla\,\cdotp (\mu H)=0.
\ea
Since $\e=1$ and $\mu=1$, we obtain using  (\ref{eqn-Maxwell}) and the basic formulae of calculus,
\ba
\Delta E=\sum_{j=1}^3 \frac {\p^2}{\p x_j^2} E
=\nabla (\nabla \,\cdotp E)-\nabla\times \nabla \times E=0-\nabla\times (ik \mu H)
=-k^2E.
\ea
This implies the Helmholtz equation $(\Delta +k^2)E=0$. Thus,   removable 
singularity results similar to those used to prove Thm. \ref{single coating with Helmholtz}
for the Helmholtz equation can be applied to Maxwell's equations to show that 
equations (\ref{eq: physical Max M}) on $N$ imply Maxwell's equations (\ref{eq: physical Max single M1 new})
first on $M_1\setminus \{0\}$ and then on all of $M_1$.
Also, analogous computations to those presented after Thm.\ \ref{single coating with Helmholtz}
for the finite energy solutions $(E,H)$ of Maxwell's equations yield that the electric
field $E$ has to satisfy the boundary condition $\nu\times E|_{\Sigma-}=0$ on the inside
of the cloaking surface. As $E$ and $H$ are in symmetric roles, it follows that
also the magnetic field has to satisfy $\nu\times H|_{\Sigma-}=0$. Summarizing,
these considerations show that the
 finite energy solutions that are also solutions in the sense of distributions, 
have outside the cloaking surface a one-to-one correspondence
to the solutions of Maxwell's equations with  the homogeneous, isotropic $\varepsilon_0$ and $\mu_0$ on $M_1$, but
inside the cloaking region  must satisfy \emph{hidden} boundary conditions at 
$\Sigma^-$.

Thm. \ref{single coating with Maxwell A} can} be interpreted by saying that the cloaking of active
objects is difficult since, with non-zero currents
present within the region to be cloaked, the idealized model
leads to non-existence of finite energy   solutions.
The theorem says that a finite energy   solution must satisfy
the hidden boundary conditions
\begin{equation}\label{eqn-pec pmc}
\nu\times \tilde E=0,\quad \nu\times \tilde H=0\quad \hbox{on } \p N_2.
\end{equation}
Unfortunately, these conditions, which correspond  physically to the
so-called perfect electrical conductor (PEC) and perfect magnetic conductor
(PMC) conditions, simultaneously, constitute
an overdetermined set of boundary conditions for Max\-well's equations on
$N_2$ (or, equivalently, on $M_2$). For cloaking passive objects, for which
$J=0$, they can be satisfied by fields which are identically zero in the
cloaked region, but
for generic  $J$, including ones arbitrarily close to 0, there is no
solution.
The perfect, ideal cloaking devices
in practice can only be approximated with a medium whose material parameters
approximate the degenerate parameters $\te$ and $\tm$.
For instance, {one} can consider metamaterials built up using
periodic structures whose effective material parameters
approximate $\te$ and $\tm$. Thus the question of when the  solutions exist
in a reasonable
sense is directly related to the question of which
approximate cloaking devices can be  built  in practice.
We note that if $E$ and $H$ solve (\ref{eq: physical Max single M1 new}),
(\ref{eq: physical Max single M2 new}), and (\ref{eq: physical Max
single M3 new})
with non-zero $b^e$ or $b^h$, then the fields
$\tilde E$ and $\tilde H$ can  be considered as solutions to a set of  
non-homogeneous
Maxwell equations
on $N$ in the sense of Definition \ref{Maxwell-def}.
\ba
\nabla\times \tilde E = ik \tilde \mu(x)  \tilde H
+\tilde K_{surf},\quad \nabla\times
              \tilde H =-ik  \tilde \e(x) \tilde E
+\tilde J+\tilde J_{surf}\quad\hbox{ on }N,
\ea
where $\tilde K_{surf}$ and $\tilde J_{surf}$ are magnetic and electric
surface
currents supported on $\Sigma$. 
{The appearance of these currents has been discussed
in \cite{GKLU1,GKLU3,Zhang2}. We note that there are many possible
choices for the currents $\tilde J_{surf}$ and $\tilde K_{surf}$.
} 
If we include a PEC lining
on $\Sigma$, that in physical terms means that we
add a thin surface made of perfectly conducting material on $\Sigma$,
the solution for the given boundary value $f$ is the one for which
{the magnetic boundary current vanish, $\tilde K_{surf}=0$ and 
the electric boundary current} $\tilde J_{surf}$
is possibly non-zero. Introducing this lining
on the cloaking surface $\Sigma$ turns out to be a remedy
for the non-existence results, and we will see that the
invisibility cloaking 
then be allowed to function as desired.

To define  the boundary
value problem corresponding to PEC lining, denote by $C^\infty_\Sigma(N)$
the space of functions $f:N\to \R$ such that
$f|_{N_1}$ and $f|_{N_2}$ are $C^\infty$ smooth up to the boundary.

\begin{definition}\label{Maxwell-def modified}
We say that  $(\tilde E, \tilde H)$ is a finite energy solution
to Maxwell's equations on $N\setminus \Sigma$ with 
perfectly conducting cloaking surface, 
\beq\label{eqn-4.1-main modified}
& &\nabla\times \tilde E = ik \tilde \mu(x)  \tilde H,\quad \nabla\times
             \tilde H =-ik  \tilde \e(x) \tilde E+\tilde J\quad
\hbox{ on }N\setminus \Sigma,\\
& &\nu\times E|_{\Sigma}=0 \nonumber
\eeq
if $ \tilde E$, $\tilde H$ are
one-forms and $\tilde D:=\tilde \e\, \tilde E$ and
    $\tilde B:=\tilde \mu\, \tilde H$ two-forms in $N$ with
$L^1(N,dx)$-coefficients satisfying conditions
(\ref{eq: Max norm1}-\ref{eq: Max norm2}),
and equations (\ref{eq: weak condition}) hold 
{for all 1-forms $\tilde e$ and $\tilde h$ on $N$
having in the Euclidian coordinates components  in $C^\infty_\Sigma(N)$,
vanishing near $\p N$, 
and satisfing $\nu\times \tilde e|_{\Sigma}=0$ from
both sides of $\Sigma.$}
\end{definition}

With such lining of $\Sigma$,  cloaking
is possible with the following result, obtained similarly to Thm.\ 5 in
\cite{GKLU1}
(cf. \cite[Thm. 2 and 3]{GKLU1}).

\begin{theorem}\label{single coating with Maxwell B}
Let $E$ and $H$ be 1-forms
with measurable coefficients
on $M\setminus \{ 0\}$ and $\tilde E$ and $\tilde H$ be
             1-forms
with measurable coefficients on $N\setminus \Sigma$
such that $\tilde E=F_*E$, $\tilde H=F_* H$.
Let $J$ and $\tilde J$ be  2-forms
with smooth coefficients on $M\setminus \{ 0\}$ and             
 $N\setminus\Sigma$, that are supported away
from $\{ 0\}$ and $\Sigma$ such that $\tilde J=F_*J$.

Then the following are equivalent:
\begin{enumerate}
\item
The 1-forms $ \tilde E$ and $ \tilde H$ on $N$ satisfy Maxwell's equations
(\ref{eqn-4.1-main modified}) 
in the sense of Definition \ref{Maxwell-def modified}.

\item
The forms $E$ and $H$ satisfy Maxwell's equations on $M$,
\beq\label{eq: physical Max single M1 new 2}
& &\nabla\times   E = ik   \mu(x)    H,\quad \nabla\times
               H =-ik    \e(x)   E+  J\quad
\hbox{ on }M_1,
\eeq
and
\beq\label{eq: physical Max single M2 new 2}
& &\nabla\times   E = ik   \mu(x)    H,\quad \nabla\times
               H =-ik    \e(x)   E+  J\quad
\hbox{ on }M_2,\\ \nonumber
& &\nu\times   E|_{\p M_2}=0.
\eeq
\end{enumerate}

\end{theorem}

The above results show that if
 we are building an approximate cloaking device with metamaterials,
effective constructions could be done in such a way that the material
approximates
a cloaking material with PEC (or PMC lining),
{which gives rise to the boundary condition on the inner part of
$\Sigma$ of the form $\nu\times   E|_{\p M_2}=0$ (or
 $\nu\times   H|_{\p M_2}=0)$.
Another physically relevant lining is the so-called SHS (soft-and-hard
surface) 
\cite{Ki1,Ki2,HLS,Li}. Mathematically, it corresponds 
to a boundary condition 
on the inner part of
$\Sigma$ of the form $E(X)=H(X)=0$, where $X$ is a tangent vector field
on
$\Sigma$. It is particularly useful for the cloaking of a 
cylinder $\{(x_1,x_2,x_3)\in \R^3: (x_1,x_2)\in D\}$, $D\subset \R^3$, 
when $X$ is the vector
$\frac{\p}{\p\theta}$ in cylindrical coordinates, see \cite{GKLU1}, \cite{GKLU3}.}
Further examples of mathematically possible boundary conditions 
on the inner part of
$\Sigma$, for a different 
notion of solution,  can be found in \cite{W3}.

The importance of the SHS lining  in the context of cylindrical cloaking is discussed
in detail  in \cite{GKLU3}.
In that case, adding a special physical surface on $\Sigma$ improves
significantly
the behavior of approximate cloaking devices;  without this kind of lining
the
fields blow up. Thus we suggest that the engineers building cloaking devices
should consider first what kind of cloak with well-defined solutions
they  would like to approximate. Indeed, building up a material
where solutions behave nicely is probably easier than building
a material with huge oscillations of the fields.

As an alternative,  one can 
 avoid the above difficulties by modifying
 the basic construction by
using a {\it  double
coating}.
Mathematically, this corresponds to using an $F=(F_1,F_2)$ with both
$F_1,F_2$ singular, which gives rise to a singular Riemannian metric
which degenerates in the same way as one approaches $\Sigma$ from both
sides.
Physically, the double coating construction corresponds to surrounding both
the
inner and outer surfaces of $\Sigma$ with appropriately matched
metamaterials, see \cite{GKLU1} for details.

\section{Electromagnetic wormholes}\label{wormholes}

We describe in this section another application of transformation optics
which consists in ``blowing" up a {curve}  rather than a point.
In \cite{GKLU2,GKLU4} a blueprint  is given for a device that
would function
as an invisible tunnel, allowing  electromagnetic
waves to propagate from one region to another, with only the ends of  the
tunnel being
visible.   Such a device, making solutions of Maxwell's equations behave as
if the topology 
of $\R^3$ has been changed to that
$\R^3 \# ({\mathbb S}^2\times {\mathbb S}^1)$, the 
connected sum of
the Euclidian space $\R^3$ and the product manifold
${\mathbb S}^2\times {\mathbb S}^1$. The connected sum
is somewhat analogous to an Einstein-Rosen wormhole \cite{ER} in general
relativity,
and so we refer to this construction
as an \emph{electromagnetic wormhole}.

We start by considering, as in Fig.\ 8,  a
3-dimensional \emph{wormhole
manifold},
$M=M_1 \cup M_2/\sim $, with components
\ba
& &M_1=\R^3\setminus (B(O,1)\cup B(P,1)),\\
& &M_2={\mathbb S}^2\times [0,1].
\ea
{Here $\sim$ corresponds to a smooth identification, \ie,
gluing, of  the boundaries
$\p M_1$ and $\p M_2$. }

An optical  device that acts as a wormhole for
electromagnetic waves at a given frequency $k$  can be constructed
by starting with a two-dimensional finite cylinder
\ba
T={\mathbb S}^1\times [0,L] \subset \R^3,
\ea
 taking its neighborhood
  $K=\{x\in \R^3:\ \dist(x,T)\le  \rho\}$,
where $\rho>0$ is small enough and defining $N=\R^3\setminus K$.
\begin{figure}[htbp]\label{schematic wormhole}
\begin{center}
\includegraphics[width=.7\linewidth]{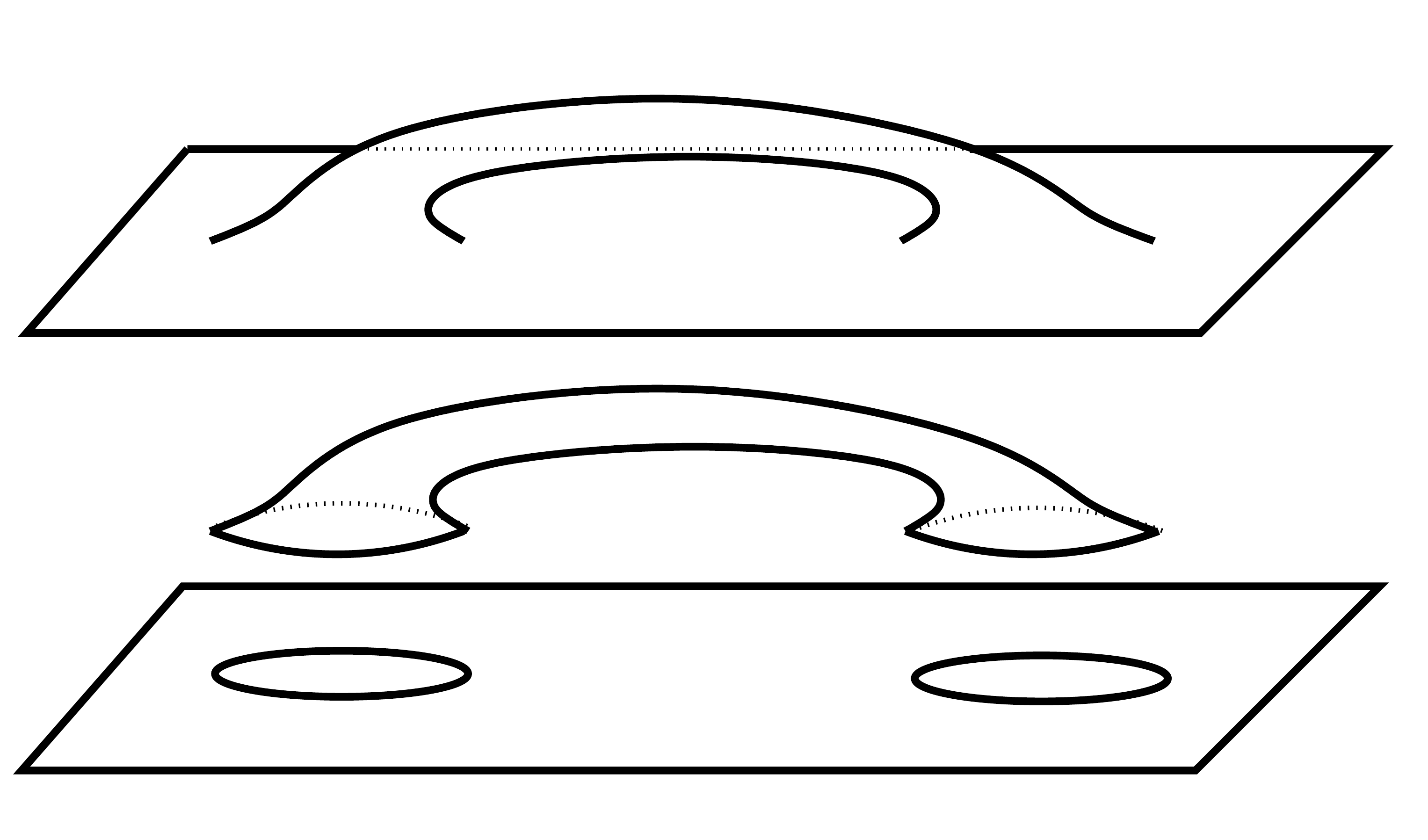}
\caption{{A two dimensional schematic figure of wormhole
construction by
gluing surfaces. Note that  the components of the artificial wormhole
construction
are three dimensional.}}
\end{center}
\end{figure}
Let us put
the SHS lining on the surface $\p K$,
 corresponding to the angular  vector field $X =\p_\theta$
in the cylindrical coordinates $(r,\theta,z)$ in $\R^3$,
 and cover $K$ with  an invisibility cloak of the single coating type.
 This material has permittivity
$\tilde \e$
and permeability $\tilde \mu$
described below, which are singular at $\p K$.
Finally, let
  \ba
U=\{x:\ \dist(x,K)>1\}\subset \R^3.
\ea
The set $U$ can be considered both as a subset of $N$,
$U \subset N(\subset \R^3)$
and of the
introduced earlier abstract wormhole manifold $M$, $U\subset
M_1$.
Let us consider the
electromagnetic measurements done in $U$, that is,
measuring fields $E$ and $H$ satisfying a radiation condition that
corresponds to an arbitrary current $J$
that is compactly supported in $U$.
 Then,
 as shown in \cite{GKLU4},  
  all electromagnetic measurements in $U\subset M$ and $U\subset N$
coincide; that is,
waves on the {\it wormhole device} $(N,\tilde \e,\tilde \mu)$ in $\R^3$
behave as if they were propagating on the abstract
{wormhole manifold} $M$.

\begin{figure}[htbp] \label{ray tracing 2}
\begin{center}
\includegraphics[width=6.4cm]{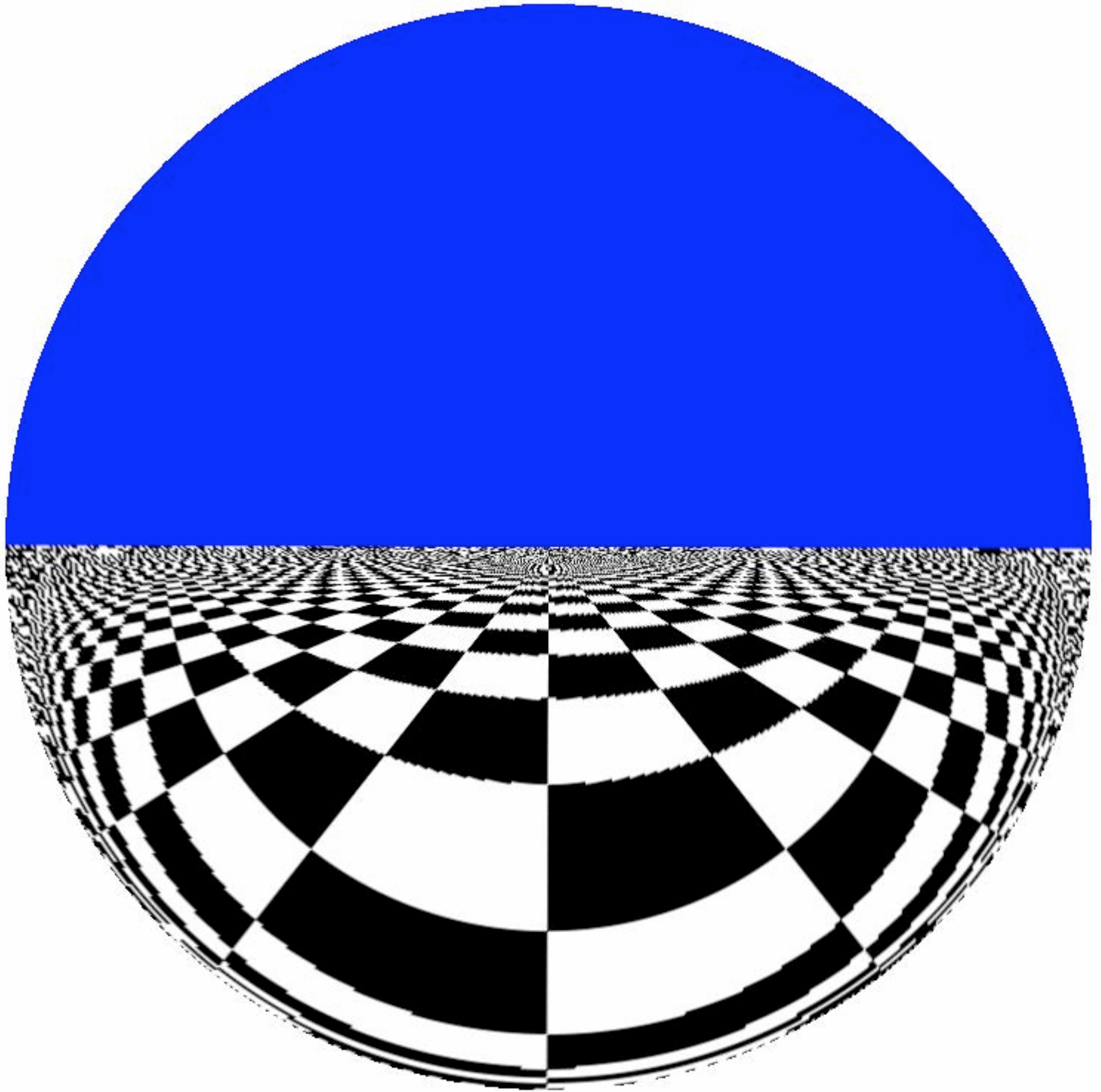}
\hspace{3mm}\includegraphics[width=6.4cm]{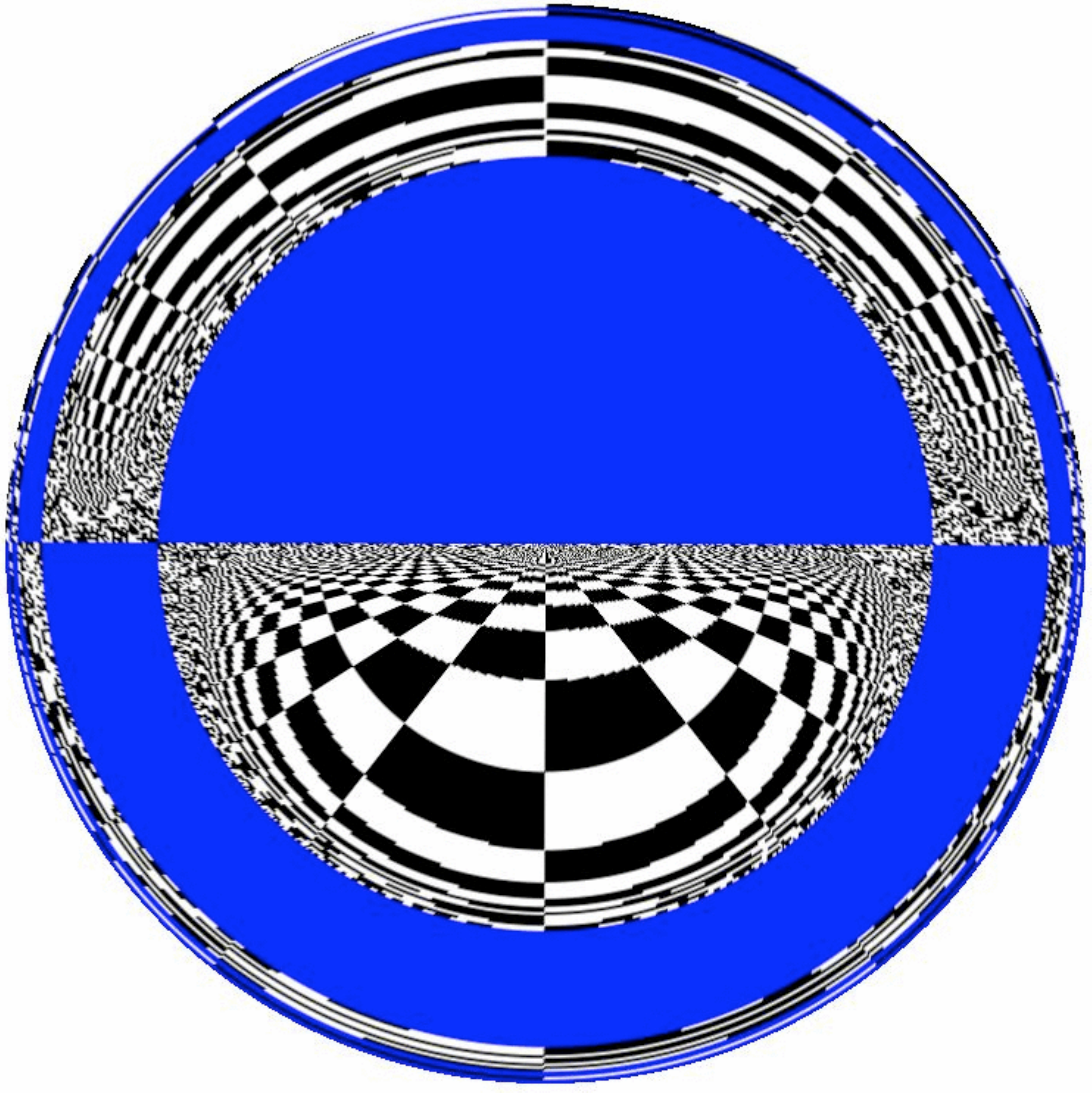}
\end{center} 
\caption{Ray tracing simulations of views through the bores of
two
wormholes.
The distant ends are above an infinite chess board under a blue sky.
On left, $L<<1$; on right, $L\approx 1$. Note
that  blue is used for  clarity; {the wormhole construction
should be considered
essentially monochromatic, for physical rather than mathematical reasons.}}
\end{figure}

In Figures 3 and 9 we give
ray-tracing simulations
in and near the wormhole. The obstacle in Fig.\ 3 is $K$,
and the metamaterial corresponding to $\tilde \e$ and $\tilde \mu$,
through which the rays travel,
is not shown.
\bigskip

We now give a more precise description of an electromagnetic wormhole.
Let us
start by making two holes in   $\R^3$,
say by removing the open unit ball $B_1=B(O,1)$, 
and also the open ball $B_2=B(P,1)$, where $P=(0,0,L)$ is a point on
the $z$-axis with $L>3$, so that $\overline{B_1}\cap\overline{B_2}=
\emptyset$.
The region so
obtained,
$M_1=\R^3\setminus (B_1\cup B_2)$, equipped with the standard Euclidian
metric $g_0$ and
a "cut"
$\g_1=\{(0,0,z): 1\le z\le L-1\}$, is
the first component $M_1$ of the wormhole manifold.

The second component of the wormhole manifold  is a $3-$dimensional
cylinder,
$M_2=\stwo\times[0,1]$, with boundary $\p M_2=(\stwo\times \{0\})\cup
(\stwo\times
\{1\}):=\stwo_3\cup\stwo_4$. We make  a "cut"
$\g_2=\{NP\}\times [0,1]$, where $NP$ denotes an arbitrary point in $\stwo$,
say the North
Pole. We initially equip
$M_2$ with the product metric, but several variations on this basic design
are possible,
having somewhat different possible applications which will be mentioned
below.

Let us glue together the boundaries $\p
M_1$
and  $\p M_2$. The glueing is done so that
we
glue the point $(0,0,1) \in  \p B({\it O}, 1)$ with the point $NP \times
\{0\}$
and the point $(0,0, L-1) \in  \p B(P, 1)$ with the point $NP \times \{1\}$.
Note that in
this  construction, $\g_1$ and $\g_2$ correspond to two nonhomotopic
curves
connecting
$(0,0,1)\sim NP\times \{0\}$ to
$(0,0,L-1)\sim NP\times \{1\}$. Moreover, $\gamma=\gamma_1\cup \gamma_2$
will be a closed
curve on $M$.

Using cylindrical coordinates, $(r,\theta,z)\mapsto (r\cos \theta ,r\sin\theta,z)$, let  
$N_2=\{(r,\theta,z):\
|r|<1,\ z\in [0,L]\}\cap N$
and $N_1=N\setminus N_2$; then consider  singular transformations
$F_j:M_j\setminus\g_j\lra \R^3,\ j=1,2$,
whose images are
$N_1,N_2$, resp., see \cite{GKLU4} for details.
{For instance,  the map $F_1$ can be chosen so that it keeps
the $\theta$-coordinate the same and maps $(z,r)$ coordinates
by $f_1:(z,r)\to (z',r')$. In the Fig.\ 10
the map $f_1$ is visualized.}

Together the maps $F_1$ and $F_2$ define a diffeomorphism
$F:M\setminus \gamma\to N$, that blows up near $\gamma$.
We define the material parameters $\tilde \e$ and $\tilde \mu$
on $N$ by setting $\tilde \e=F_*\e$ and  $\tilde \mu=F_*\mu$.
These material parameters (having freedom in choosing the map $F$)
give blueprints for how a wormhole device could be constructed
in the physical space $\R^3$.

\begin{figure}[htbp]\label{Siam Figure}
\begin{center}
\psfrag{1}{}
\psfrag{2}{}
\psfrag{3}{}
\psfrag{4}{}
\psfrag{5}{}
\psfrag{6}{}
\psfrag{7}{}
\psfrag{8}{}
\psfrag{9}{\hspace{-1mm}$f_1$}
\psfrag{A}{$\vspace{-2mm}Q$}
\psfrag{B}{$R$}
\psfrag{C}{$P$}

\includegraphics[width=14cm]{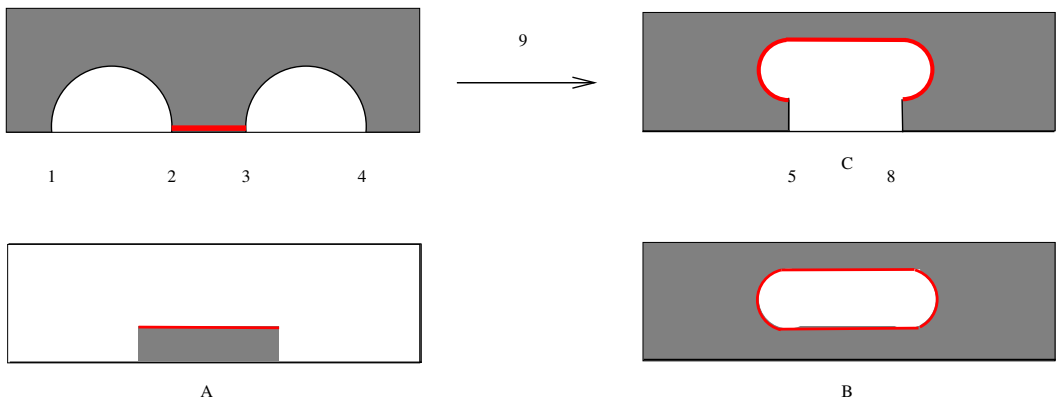} \label{pic 5}
\end{center}
\vspace{-1cm} 
\caption {{\bf Above:} {A schematic figure of
$f_1$,
representing
$F_1$,  in the $(z,r)$ plane.} Its image
$P$ corresponds to $N_1$ in $(z,r)$ coordinates.
{\bf Below:} {The sets $Q$ and $R$ correspond to $N_2$ and $N$.
In
the figure, $R=Q\cup P$ which corresponds to $N=N_1\cup N_2$ in $\R^3$.}}
\end{figure}

\begin{figure}[htbp] \label{rays return back}
\begin{center}
\includegraphics[width=.9\linewidth]{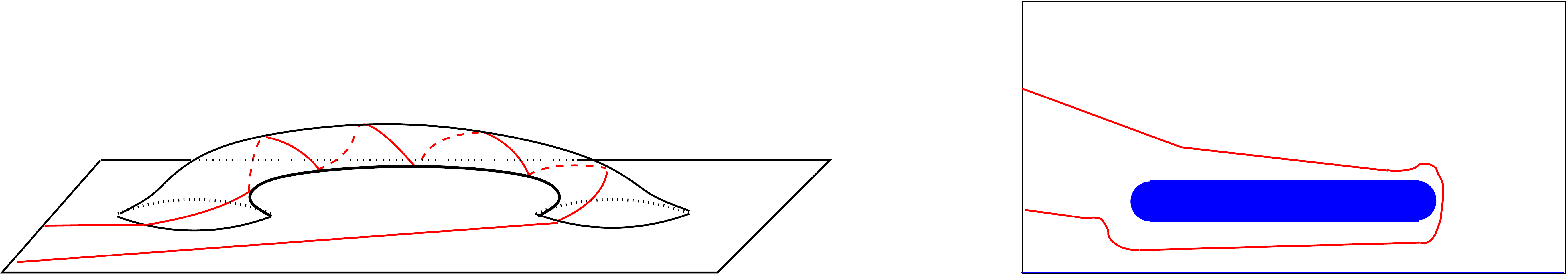}
\end{center}
\caption{{Schematic figure. {\bf Left:} Some rays enter the wormhole and come
out from
the other end so that they return near where
the
ray entered to the wormhole. {\bf Right:} The corresponding ray
in the complement $N$ of the obstacle $K$ shown in the
$(z,r)$ coordinates. Note that there are also closed light rays.}
}
\end{figure}

{Possible applications of  electromagnetic wormholes (with varying
degrees of likelihood
of realization!),  when the   metamaterials technology
has sufficiently progressed,}
include invisible optical cables, 3D video displays,
scopes for
MRI-assisted medical procedures, and beam collimation. For the last two, one
needs to
modify the design by changing the metric $g_2$ on $M_2=\stwo\times [0,1]$.
By flattening
the metric on $\stwo$ so that {the antipodal point $SP$ (the south
pole) to
$NP$} has a neighborhood on which the metric is
Euclidian, the
axis of the tunnel $N_2$ will have a tubular neighborhood on which $\e,\mu$
are constant
isotropic and hence can be allowed to be empty space, allowing for passage
of instruments.
On the other hand, if we use a  warped product metric on $M_2$,
corresponding to
$\stwo\times\{z\}$ having the metric of the sphere of radius $r(z)$ for an
appropriately
chosen function $r:[0,1]\lra\R_+$, then only rays that travel through $N_2$
almost parallel to
the axis can pass all the way through, with others being returned to the end
from which
they entered.

\begin{remark}
{\rm
Along similar lines, we can produce another interesting class of
devices,
made possible with
the use of metamaterials, which behave as if the topology of $\R^3$ is
altered. Let $M_1= \R^3 \setminus B(0,1)$ endowed with the
Euclidian
metric $g$ and $M_2$ be
a copy of $M_1$. Let $M$ be the manifold obtained by glueing the
boundaries $\p M_1$ and $\p M_2$ together. Then $M$ can be considered as
a $C^\infty$ smooth manifold with Lipschitz smooth metric.
Let $N_1=M_2$ and $F_1:M_1\to N_1$ be the identity map,
$N_2=B(0,1)\setminus \overline B(0,\rho)$ with $\rho\geq 0$, and finally
$N=\R^3\setminus \overline B(0,\rho)$.
Let $F_2:M_2\to N_2$ be the map $F_2(x)=(\rho+(1-\rho)|x|^{-1})|x|^{-1}x$.
Together the maps $F_1$ and $F_2$ define a map that can be extended to a
Lipschitz smooth diffeomorphism
$F:M\to N$. As before, we define on $N$ the metric $\tilde g=F_*g$,
and the permittivity $\tilde \e$ and permeability $\tilde \mu$
according to formula (\ref{def: tilde e and mu}).
As $N=\R^3\#\R^3$, we can consider the $(N,\tilde \e,\tilde \mu)$
as a {\it parallel universe} device on which the electromagnetic waves on
$\R^3\#\R^3$
can be simulated. It is particularly interesting to consider the high
frequency case when the ray-tracing leads to  physically interesting
considerations. Light rays correspond to the locally shortest curves on
$N$, so
{all rays emanating from $N_1$ that do not hit $\p N_1$ tangentially
then enter
$N_2$.
Thus the light rays in $N_1$ that hit  $\p N_1$ non-tangentially
 change the sheet $N_1$ to $N_2$. From the point of view
of an observer in $N_1$, the rays are  absorbed
by the device. Thus on the level of ray-tracing the device is
a perfectly black body, or a perfect absorber.
Similarly analyzing the quasi-classical solutions,
the energy, corresponding to the non-tangential directions is
absorbed,  up to the first order of magnitude, by the device.}
Other, 
metamaterial-based constructions of a perfect absorber have been considered in \cite{LSMSP}.
We note that in our considerations the energy is not in reality absorbed
as there is no dissipation in the device, and thus  the energy is in fact
trapped inside the device, which naturally causes difficulties in
practical implementation. On the level of the ray tracing similar
considerations
using multiple sheets have been considered before in \cite{LeT}.
}
\end{remark}

\section{A general framework:\\
singular  transformation optics}\label{sec-transf opt}

Having seen how cloaking based on blowing up a point or blowing up a line
can be rigorously analyzed, we now want to explore how more general optical
devices can be described using the transformation rules satisfied by $n,
(\rho,\lambda),\e$ and
$\mu$. This point of view has been advocated by J. Pendry and his
collaborators,
and given the name \emph{transformation optics} \cite{ward}.
As
discussed earlier, under a nonsingular changes of variables $F$, there is a
one-to-one correspondence between solutions $\tu$ of the relevant
equations for
the transformed medium  and solutions
$u=\tu\circ F$ of the original medium. However, when $F$ is {\it singular}
at
some points, as is the case for cloaking {and the wormhole}, we
have shown how greater care
needs to be taken, not just for the sake of mathematical rigour, but to
improve the cloaking effect for more physically realistic approximations to
the ideal material parameters. Cloaking {and the wormhole} can
be considered as merely
starting points for what might be termed
\emph{singular transformation optics}, which, combined with the rapidly
developing technology of metamaterials, opens up entirely new possibilities
for designing  devices having novel effects on acoustic or electromagnetic
wave propagation. Other singular transformation designs  in 2D that rotate
waves within the cloak \cite{Ch1}, concentrate waves \cite{Ra1} or act as
beam splitters \cite{Ra2}  have been proposed. Analogies with phenomena in
general
relativity have been {proposed in \cite{LeP} as a source of
inspiration for designs.}

We  formulate a general approach to the precise description of the
ideal material parameters in a singular transformation optics device,
$N\subset\R^3$, and state a ``metatheorem", analogous to the results we have
seen above, which should, in considerable generality, give an exact
description of the electromagnetic waves propagating through such a device.
However, we wish to stress that, as for cloaking \cite{GKLU1} and the
wormhole \cite{GKLU2,GKLU4}, actually proving this ``result" in
particular cases of
interest, and determining the hidden boundary conditions, may be decidedly
nontrivial.

{A general framework for considering ideal mathematical
descriptions of such designs  is as follows: Define}
a \emph{singular transformation optics (STO) design} as a triplet
$(\M,\N,\F)$, consisting of:

\num{i} An STO \emph{manifold}, $\M=(M,g,\g)$, where
$M=(M_1,\dots,M_k)$, the disjoint union of $n$-dimensional Riemannian
manifolds $(M_j,g_j)$, with or without boundary, and
(possibly empty)  submanifolds $\gamma_j\subset
\hbox{int } M_j$, with $\dim \gamma_j\le n-2$;

\num{ii} An STO \emph{device}, $\N=(N,\Sigma)$, where
$N=\bigcup_{j=1}^k N_j\subset \R^n$ and
$\Sigma=\bigcup_{j=1}^k \Sigma_j$, with $\Sigma_j$ a (possibly empty)
hypersurface in $N_j$; and

\num{iii} A \emph{singular transformation} $\F=(F_1,\dots,F_k)$, with
each
\linebreak$F_j:M_j\setminus\g_j\lra N_j\setminus\Sigma_j$ a
diffeomorphism.
\bigskip

Note that $N$ is then equipped with a singular Riemannian
metric $\tg$, with $\tg|_{N_j}=(F_j)_*(g_j)$, in general degenerate on
$\Sigma_j$. Reasonable conditions need to be placed on the Jacobians $DF_j$
as
one approached $\g_j$ so that the $\tg_j$ have the appropriate degeneracy,
cf.
\cite[~Thm.3]{GLU3}.

In the context of the conductivity or Helmholtz equations, we can then
compare solutions
$u$ on $\M$ and $\tu$ on $\N$, while for Maxwell  we can
compare fields $(E,H)$ on $\M$ (with $\e$ and $\mu$ being the Hodge-star operators
corresponding to the metric $g$) and $(\tE,\tH)$ on $\N$.
For simplicity, below we  refer  to the fields as just $u$.
\bigskip

\noindent{\bf Principle of Singular Transformation Optics,  or   ``A
Metatheorem about Metamaterials":}  {\it If $(\M,\N,\F)$ is an STO
triplet, there is a 1-1 correspondence,
given by
$u=\tu\circ
\F$, \ie, $u|_{M_j}=(\tu|_{N_j})\circ F_j$,
between finite energy   solutions  $\tu$ to the equation(s) on
$\N$, with
source terms $\tilde{f}$ supported on $\N\setminus \Sigma$,
and finite energy solutions $u$ on $\M$,
with source terms $f=\tilde{f}\circ\F$,
satisfying certain ``hidden" boundary
conditions on
$\p M=\bigcup_{j=1}^k\, \p M_j$.}

\section{Isotropic Transformation Optics}\label{sec-ito}

The design of transformation optics (TO)  devices, based on the
transformation rule (\ref{eqn1.6}), invariably leads to anisotropic material
parameters.
Furthermore, in
singular TO designs,  such as cloaks,
field rotators \cite{Ch1}, wormholes \cite{GKLU2,GKLU4}, beam-splitters
\cite{Ra2}, or any of those arising from the considerations of the previous
section, the material parameters are singular, with one or
more eigenvalues going to 0 or $\infty$ at some points.

While raising interesting mathematical  issues, such singular, anisotropic
parameters are difficult to physically implement. The area of
metamaterials is developing rapidly, but fabrication of highly
anisotropic and (nearly) singular materials at frequencies of interest
will clearly remain a challenge for some time. Yet another constraint
on the realization of theoretically perfect ( or \emph{ideal} in the
physics nomenclature) TO designs is \emph{discretization}: the
metamaterial cells have positive diameter and any physical construction
can represent at best a discrete sampling of the ideal parameters.

There is a way around these difficulties. At the price of losing the
theoretically perfect effect on wave propagation that ideal TO designs
provide, one can gain the decided advantages of being able to use discrete
arrays of metamaterial cells with isotropic and nonsingular material
parameters. The procedure used in going from the anisotropic, singular
ideal  parameters to the isotropic, nonsingular, discretized 
parameters involves techniques from the analysis of variational problems,
homogenization and spectral theory. We refer to the resulting designs 
as arising from  \emph{isotropic transformation optics}. How this is carried
out is sketched below in the context of cloaking; more details and
applications  can be found in \cite{GKLU6,GKLU7,GKLU8}.

The initial step is to truncate  ideal cloaking material parameters,
yielding a nonsingular, but still anisotropic, approximate cloak; similar
constructions have been used previously in the analysis of
cloaking \cite{RYNQ,GKLU3,KSVW, CWZK}. This approximate cloak is then
itself approximated by  nonsingular, isotropic parameters. The first
approximation is justified using the notions of $\Gamma$- and
$G$-convergence from variational analysis \cite{A,dM}, while the second
uses more recent ideas from \cite{Allaire,Allaire2,Cherka}.

We start with the ideal spherical cloak for the acoustic wave equation.
For technical reasons, we modify slightly  the cloaking conductivity
(\ref{eq: cond})
by setting it equal to $2\delta^{jk}$ on $B(0,1)$, and relabel it as
$\sigma$
for simplicity.
Recall that $\sigma$  corresponds to a singular Riemannian
metric $g_{jk}$
that is related to $\sigma^{ij}$ by
\beq \label{determinant}
\sigma^{ij}(x)= |g(x)|^{1/2} g^{ij}(x), \quad |g|
= \left( \hbox{det}[\sigma^{ij}] \right)^2
\eeq
where $[g^{jk}(x)]$ is the inverse matrix of $[g_{jk}(x)]$ and
  $|g(x)|= \hbox{det}[g_{jk}(x)]$.
The resulting Helmholtz equation, with a source term $p$,
\beq\label{case 1}
& &\sum_{j,k=1}^3 |g(x)|^{-1/2}\frac \p{\p x^j}(|g(x)|^{1/2}g^{jk}(x)
\frac \p{\p x^k} u)+\omega^2 u= p\quad\hbox{on }N,\\
& &u|_{\p N}=f, \nonumber
\eeq
can then be reinterpreted by thinking of $\sigma$ as a \emph{mass} tensor
(which
{indeed has the same transformation law
as  conductivity under coordinate diffeomorphisms} )
and $|g|^\frac12$ as a \emph{bulk
modulus} parameter; (\ref{case 1}) then becomes an acoustic wave
equation at frequency $\omega$
{with the new source  $p |g|^{1/2}$},
\beq\label{case 3 B}
& &\left(\nabla \cdotp\sigma \nabla +\omega^2|g|^{\frac12}
\right)u= p(x) |g|^{\frac12} \quad\hbox{on }N,\\
& &u|_{\p N}=f. \nonumber
\eeq
This is the form of the acoustic wave equation considered in
\cite{Ch3,Cu,GKLU5}. (See also \cite{CuSc} for $d=2$, and \cite{Norris}
for cloaking with both mass and bulk modulus anisotropic.)
To consider equation (\ref{case 3 B}) rigorously,
we assume that the source $p$ is supported away from the surface $\Sigma$.
Then the finite energy solutions $u$ of equation (\ref{case 3 B})
are defined analogously to Def. \ref{energysolution}.
Note that the function $|g|^{1/2}$ appearing in  (\ref{case 3 B}) 
is bounded from above.

Now truncate this ideal acoustic cloak: for each $1<R<2$, let
$\rho=2(R-1)$ and define
$F_R:\R^3 \setminus B(0,\rho)\to \R^3 \setminus B(0,R)$ by
 \ba
x:=F_R(y)=\left\{\begin{array}{cl} y,&\hbox{for } |y|>2,\\
\left(1+\frac {|y|}2\right)\frac{y}{|y|},&\hbox{for }\rho<|y|\leq 2.
\end{array}\right.
\ea
We define the  corresponding approximate conductivity, $\sigma_R$ as
\beq \label{R-ideal}
\sigma^{jk}_R(x)=\left\{\begin{array}{cl }\sigma^{jk}(x)
&\hbox{for } |x|>R,\\
2 \delta^{jk},&\hbox{for }|x|\leq R, \end{array}\right.
\eeq
{where $\sigma^{jk}$ is the same  as in the first formula in
(\ref{eq: cond}) or, in spherical coordinates, (\ref{eqn-sing tensor 2}).}
Note that then $\sigma^{jk}(x)=
\left(\left(F_{R}\right)_*\sigma_0\right)^{jk}(x)$ for $|x|>R$, where
$\sigma_0\equiv 1$ is the homogeneous, isotropic  mass density
tensor.  Observe that, for each $R>1$,  $\sigma_R$ is nonsingular,
i.e., is bounded from above and below, but with the lower bound going to $0$
as $R \searrow 1$. Now define 
\beq \label{R-equation}
g_R(x)=\det(\sigma_R(x))^2=
\left\{\begin{array}{cl}
64|x|^{-4}(|x|-1)^4&\hbox{for } R<|x|<2,\\
64,&\hbox{for }|x|\leq R, \end{array}\right.
\eeq
cf. (\ref{determinant}). Similar to (\ref{case 3 B}),
consider the solutions of
\beq\label{eq: Helmholtz with R}
(\nabla \cdotp\sigma_R \nabla +\omega^2 g_R^{1/2})u_R
&=&g_R^{1/2} p\quad\hbox{in }N\\ \nonumber
u_R|_{\p N}&=&f.
\eeq

As in Thm. \ref{single coating with Helmholtz},  
by considering $F_R$ as a transformations of coordinates one sees that
\ba
u_R(x)=\left\{\begin{array}{cl} v_R^+(F_R^{-1}(x)),&\hbox{for } R<|x|<2,\\
v_R^-(x),&\hbox{for } |y|\leq R,\end{array}\right.
\ea
with $v_R^{\pm}$ satisfying
\ba
(\Delta +\omega^2)v_R^+(y)&=&p(F_R(y)) \quad \hbox{in
}\rho<|y|<2,\\ v_R^+|_{\p B(0,2)}&=&f, \ea
and
\beq \label{extra-equation}
(\nabla^2+4\omega^2)v_R^-(y)&=&4p(y), \quad \hbox{in }|y|<R.
\eeq
Since $\sigma_R$ and $g_R$ are  nonsingular everywhere,
we have
the standard transmission conditions
on  $\Sigma_R:=\{x:\ |x|=R\}$,
\beq\label{trans a1}
& & u_R|_{\Sigma_R+}=u_R|_{\Sigma_R-},\\
\label{trans a2}
& &
e_r\cdotp \sigma_R \nabla u_R|_{\Sigma_R+}=
 e_r\cdotp \sigma_R \nabla u_R|_{\Sigma_R-},
\eeq
where $e_r$ is the radial unit vector and $\pm$ indicates when the
trace on $\Sigma_R$ is computed as the limit $r\to R^\pm$.

The resulting solutions, say for either no source, or for $p$ supported at
the origin,
can be analyzed using spherical harmonics, and one can show that the waves
$v$ for the ideal cloak are the limits of the waves for the approximate
cloaks, with the Neumann boundary condition in (\ref{eq on M2}) for the
ideal cloak emerging from the behavior of the waves $v_R^\pm$ for the
truncated cloaks. 
This can be seen using  spherical coordinates and observing that the trace
of the
radial component of conductivity from outside, $\sigma_R^{rr}|_{\Sigma_R+}$,
 goes to zero as $R\to 1$ but the trace $\sigma_R^{rr}|_{\Sigma_R+}$ from
inside stays bounded from below. Using this, we can see that the
transmission
condition (\ref{trans a2}) explains  the appearance of the Neumann boundary
condition
on the inside of the cloaking surface.

To consider general conductivities, {we recall that for a conductivity
$\gamma^{jk}(x)$ that is bounded both from above and below, 
the solution of the boundary value problem 
(\ref{eqn1.1}) in $N$ is the unique 
minimizer of the quadratic form
\beq\label{minimization problem}
{\mathcal Q}_\gamma (v)=\int_N \gamma\nabla v\,\cdotp \nabla v\,dx
\eeq
over the functions  $v\in H^1(N)$ satisfying  the boundary condition $v|_{\p N}=f$.

We use the above to consider the truncated conductivities $\sigma_R$.} Note that at each point $x\in N$
the non-negative matrix $\sigma_R(x)$ is a decreasing function of $R$. Thus
the quadratic forms $v\mapsto {\mathcal Q}_{\sigma_R}(v)$ 
are pointwise decreasing. As the minimizer $v$ of the quadratic form $
{\mathcal Q}_{\sigma_R}(v)+\bra h,v\cet_{L^2}$
with the condition $v|_{\p N}=f$ is the solution of the equation
\ba
\nabla\cdotp \sigma_R\nabla v=h,\quad v|_{\p N}=f,
\ea
we can use
methods from  variational analysis, in particular $\Gamma$-convergence (see,
\eg,  \cite{dM}) to consider solutions of equation (\ref{eq: Helmholtz with
R}).
Using that,
it is possible to
show that the solutions $u_R$ of the approximate equations (\ref{eq:
Helmholtz with R})
converge to the solution $u$ of (\ref{case 3 B}) for the general sources $p$
not supported on $\Sigma$ in the case when $\omega^2$ is not an eigenvalue
of the equation  (\ref{case 3 B}).

Next, we approximate the nonsingular but anisotropic conductivity $\sigma_R$
with isotropic tensors.
One can show that there exist nonsingular, isotropic conductivities
$\gamma_n$
 such that the solutions of
\beq\label{isotropic Helmoltz}
& &\left(g_R(x)^{-1/2}\nabla \cdotp\gamma_n(x) \nabla +\omega^2 \right)u_n=
p\quad\hbox{on }N,\\ & &u_n|_{\p N}=f, \nonumber
\eeq
tend to the solution of (\ref{eq: Helmholtz with R}) as $n\to\infty$.
This is obtained by considering isotropic conductivities
$\gamma_n(x)=h_n(|x|)$
depending only on radial variable $r=|x|$, where $h_n$ oscillates between large
and
small values. Physically, this corresponds to layered spherical shells
having high
and low conductivities. As the oscillation of $h_n$ increases, these
spherical
shells approximate an anisotropic medium where the conductivity has much
lower
value in the  radial direction than in the angular variables. Roughly
speaking,
currents can easily flow in the angular directions on the highly conducting
spherical shells, but the currents flowing in the radial direction must
cross both
the low and high conductivity shells. Rigorous analysis based on 
homogenization theory
\cite{Allaire, Cherka} is used for $\omega^2 \in \R_-$, and one can see
that, with
appropriately chosen isotropic conductivities
$\gamma_n$,
the solutions $u_n$ converge to the limit $u_R$. These considerations can be
extended to all
$\omega^2 \in \C \setminus {\it D}$, where ${\it D} \subset \R_-$ is
a discrete set, by  spectral-theoretic  methods, \cite{Ka}.
More details can be found in \cite{GKLU6,GKLU8}.

Summarizing, considering equations (\ref{isotropic Helmoltz}) with
appropriately
chosen smooth isotropic conductivities $\gamma_n$ and bulk moduli $g_R$
and 
letting $n \to \infty$ with $R=R(n) \to 1$, we obtain Helmholtz equations
with isotropic and non-singular mass and bulk modulus,
whose solutions converge to the solution of the ideal invisibility
cloak (\ref{case 3 B}).

A particularly interesting application of the above construction is to
quantum mechanics.
Zhang, et al. \cite{Zhang} described an anisotropic mass tensor $\hat{m}$
and a potential $V$ which together act as a cloak for matter waves, \ie,
solutions of the corresponding anisotropic Schr\"odinger equation. This
ideal quantum cloak is the result of applying the same singular transform
$F$ as used for conductivity, Helmholtz and Maxwell, and applying it to
the Schr\"odinger equation with mass tensor $\hat{m}_0=\delta^{jk},\,
V_0\equiv 0$. Due to the anisotropy of $\hat{m}$, and the singularity of
both $\hat{m}$ and $V$, physical realization would be quite challenging.
However, using the approximate acoustic cloak, one can describe an
approximate quantum cloak that should be much easier to physically realize.
An analogue of the reduction (\ref{eq:2.1}),  (\ref{eq:2.2}) of the
isotropic conductivity equation  to a Schr\"odinger equation can be
carried out for
the acoustic equation. Letting $E=\omega^2$,  $\psi_n(x)=\gamma_n^{1/2}(x)
u_n(x)$, and

\beq
\label{effective-potential}
V_n^E(x):\,=\gamma_n ^{-1/2}\nabla^2\gamma_n ^{1/2} (x)-E \gamma_n^{-1}
g_{n}^{1/2}+E,
\eeq
one computes that $\psi_n$ satisfies
{the Schr\"odinger equation}
\ba
(-\Delta+V_n^E)\psi_n=E\psi_n \quad \hbox{in }\,\, N.
\ea
Furthermore, the family $\{V_n^E\}$ acts an approximate cloak at energy $E$:

\begin{theorem} \emph{Approximate quantum cloaking.}
Let $W$ be a potential $W\in L^\infty(B(0,1))$, and $E \in \R$ not be  a
Dirichlet eigenvalue of $-\Delta$ {on $N=B(0,2)$},
nor a Neumann eigenvalue of $-\Delta+W$ on $B(0,1)$.
 The DN operators  at $\p N$  for the Schr\"odinger operators
corresponding to the  potentials
$W+V_n^E$
converge to the DN operator corresponding to free space, that is,
\ba
\lim_{n\to \infty} \Lambda_{W+V_n^E}(E) f= \Lambda_0(E) f
\ea
in $L^2(\p N)$ for any smooth $f$ on $\p N$.

The convergence of the DN operators also implies convergence of
the scattering amplitudes \cite{Ber}: $
\lim_{n\to\infty} a_{W+V_n^E}(E,\theta',\theta)=a_0(E,\theta',\theta)$.
\end{theorem}

(Note that this is not a consequence of  standard results from 
perturbation theory, since the
$V_n^E$ do not tend to $0$ as $n \to \infty$. Rather, as $n \to \infty$,
the $V_n^E$ become highly oscillatory near $\Sigma$ and $\sup_x
|V_n^E(x)|\to\infty$ as $n\to\infty$.)

\begin{figure}[htbp]\label{almost trapped}
\begin{center}
\centerline{
\includegraphics[width=.80\linewidth]{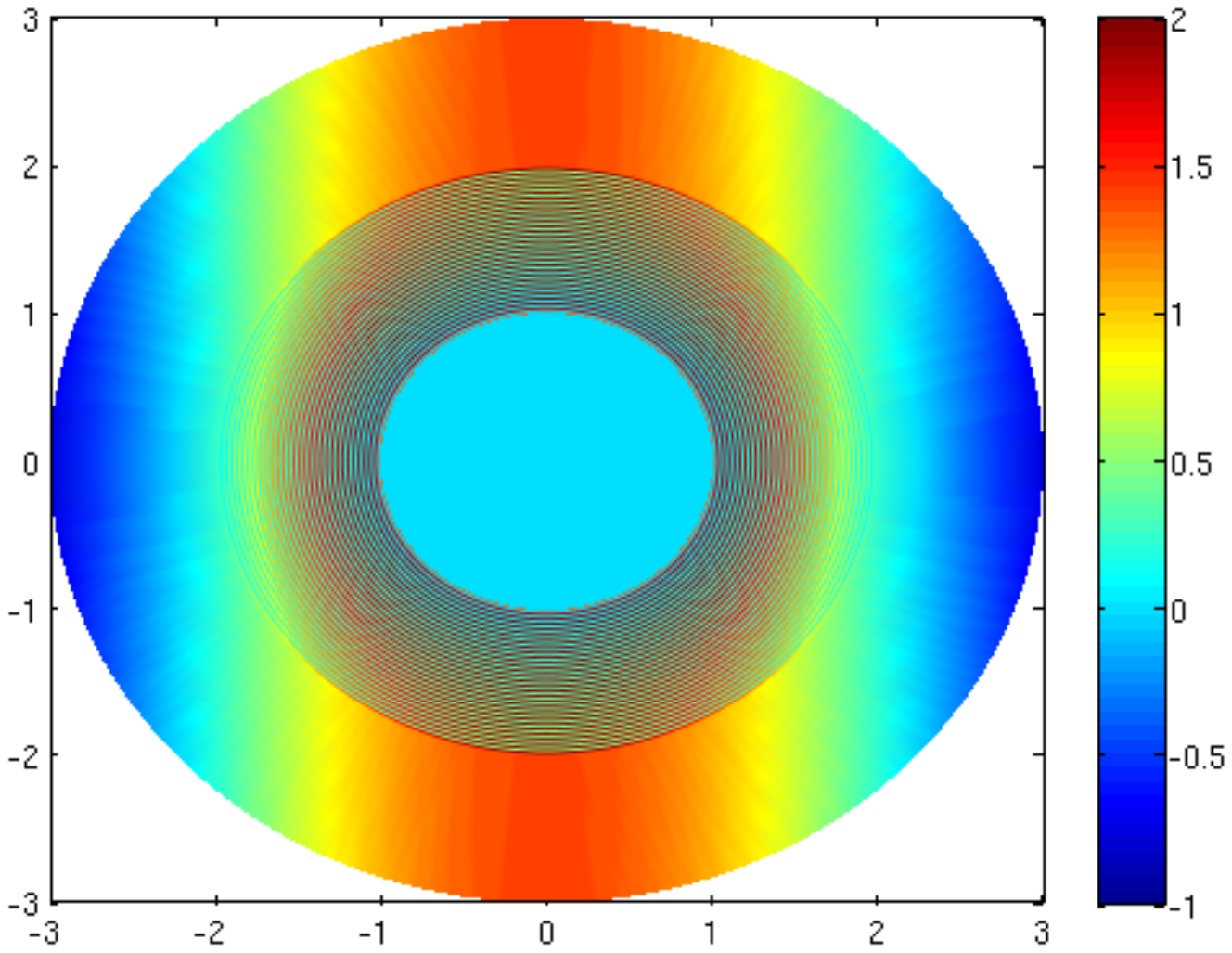}
\hspace{-3cm}
\includegraphics[width=.80\linewidth]{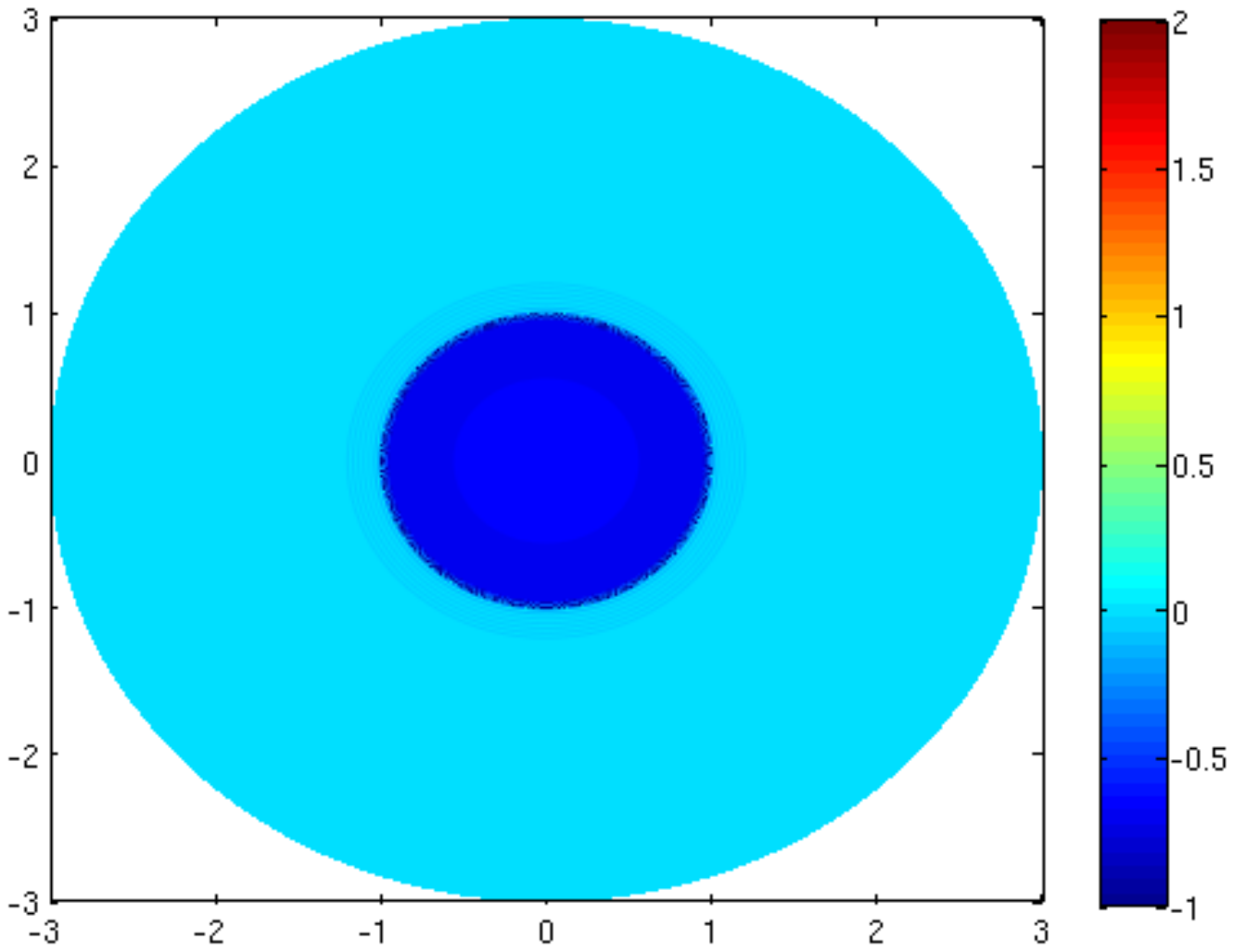}
}
\end{center}
\caption{ {\bf Left:} $E$ not a Neumann eigenvalue; \emph{approximate quantum cloak}. Matter wave passes almost unaltered.  {\bf Right:} $E$ a Neumann eigenvalue; potential supports {\it almost trapped state.} }
\end{figure}

On the other hand, when $E$ \emph{is} a Neumann eigenvalue of $-\Delta +W$ on $B(0,1)$, then $V^E_n$ supports \emph{almost trapped states}, which correspond to matter waves (\ie, quantum mechanical particles) which reside in $B(0,1)$ with high probability. See Fig.\ 12 and \cite{GKLU7} for more details and applications.

\begin{remark}
\label{enforced}
{\rm Parameter distribution similar to (\ref{R-ideal}) have been studied in the physics literature in the context of realistically  achievable layouts of
metamaterials  approximating an ideal cloak. Using other, apparently only slightly different designs,
one obtains in the limit other cloaking devices with enforced boundary
conditions on the
inside of $\Sigma$;  see \cite{GKLU6},  where approximate cloaks are
specified which give rise instead  to the Robin boundary condition.}
\end{remark}

\section{Further developments and open problems}\label{sec-add}

The literature on metamaterials, cloaking, and 
transformation optics is growing rapidly .
We  briefly describe here only a few  recent developments
and remaining challenges. See \cite{NJP} for a variety of perspectives.

\num{a}  Although the first description \cite{GLU2,GLU3} of the cloaking phenomenon was in the context of electrostatics, no proposals of electrostatic metamaterials
that might be used to physically implement these examples  have
been made to date.   A proposal for metamaterials
suitable  for magnetostatics (cloaking for which is of course
mathematically identical to electrostatics), and magnetism at very low
frequencies, is in \cite{wood}. Since \cite{Sc}, there has been a push to obtain cloaking at higher frequencies, with the visual part of the electromagnetic spectrum an obvious goal. 
Progress has been reported in \cite{Cai,smol,Liu,Shv}. However,  broadband  visual cloaking seems at this point to be far off. It should also be pointed out that
serious skepticism concerning the practical
advantages of transformation optics based cloaking
over earlier techniques for reducing scattering
has been expressed  \cite{Ki3}.

\num{b} Other boundary conditions at the cloaking surface, 
analyzed in the time domain,
based on Von Neumann's theory of self-adjoint extensions and using
a different notion of solution  than that considered here,
have been  studied in \cite{W1,W2,W3}. See also \cite{Ya}.

\num{c}  For simplicity, in cloaking we have mainly considered singular transformations which are affine linear in $r$. (See, however, Thm.\  \ref{main B}.)
In situations where the measurements are made further from
cloaked object, \cite{Cai2} introduced, for
spherical cloaking, transformations nonlinear
in the radial variable in order to give better impedance matching
with the surrounding media, and this was further explored
for cylindrical cloaking in \cite{YRQ}.

\num{d}  Effective medium theory for metamaterials is in its
early development, and seems to be particularly
difficult for materials assembled from periodic or almost-periodic arrays of
small cells whose properties
are based on resonance effects.
A physical (although mathematically nonrigorous) analysis of this kind of
media  is in \cite{SP}, which makes implicit assumptions about the
smoothness of the fields which are violated when the fields experience the
blow up demonstrated in \cite{RYNQ,GKLU3}.
Some recent work on homogenization in this context is in \cite{Ko}.
However, further  efforts in this directions are needed.

\num{e}
Existing theories of cloaking deal predominantly with
non-relativistic media; see, however, \cite{LeP}.
It seems that  developing a theory compatible with the
relativistic framework would be important. Similarly, transformation optics in
the context of nonlinear media seems likely to become significant
as metamaterial technology develops.

\num{f} At the cloaking surface $\Sigma$ the  cloaking metric $\tilde{g}$ on $N_1$ has a conical singularity in the sense of geometric scattering theory. It would be interesting
to understand the relationship between cloaking and other transformation optics constructions on the one hand and   geometric scattering on the other.

\bibliographystyle{amsalpha}

\end{document}